\documentstyle[psfig]{l-aa}
\newcommand{\beq}{\begin{equation}}
\newcommand{\beqa}{\begin{eqnarray}}
\newcommand{\eeq}{\end{equation}}
\newcommand{\eeqa}{\end{eqnarray}}
\newcommand{\figun}{
\begin{figure}[ht!]
\centering%
\psfig{file=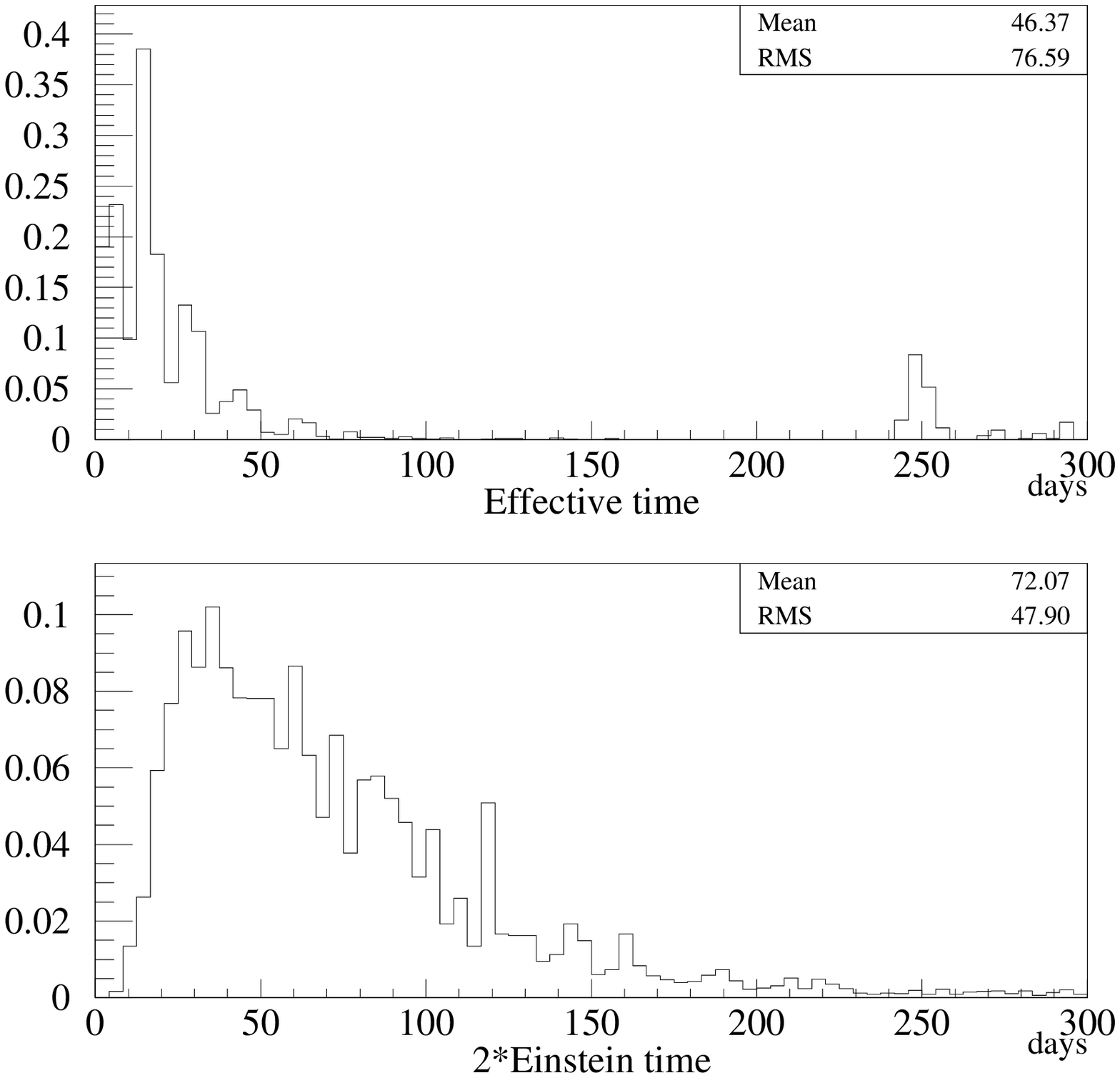,width=.48\textwidth}%
\caption[]{{\em Simulated} distributions of the effective duration%
$t_{\mathrm{eff}}$ and twice the Einstein time $t_E$, for ``detected%
events''.\label{fig01}}%
\end{figure}
}
\newcommand{\figdeux}{
\begin{figure}[ht!]%
\centering%
\psfig{file=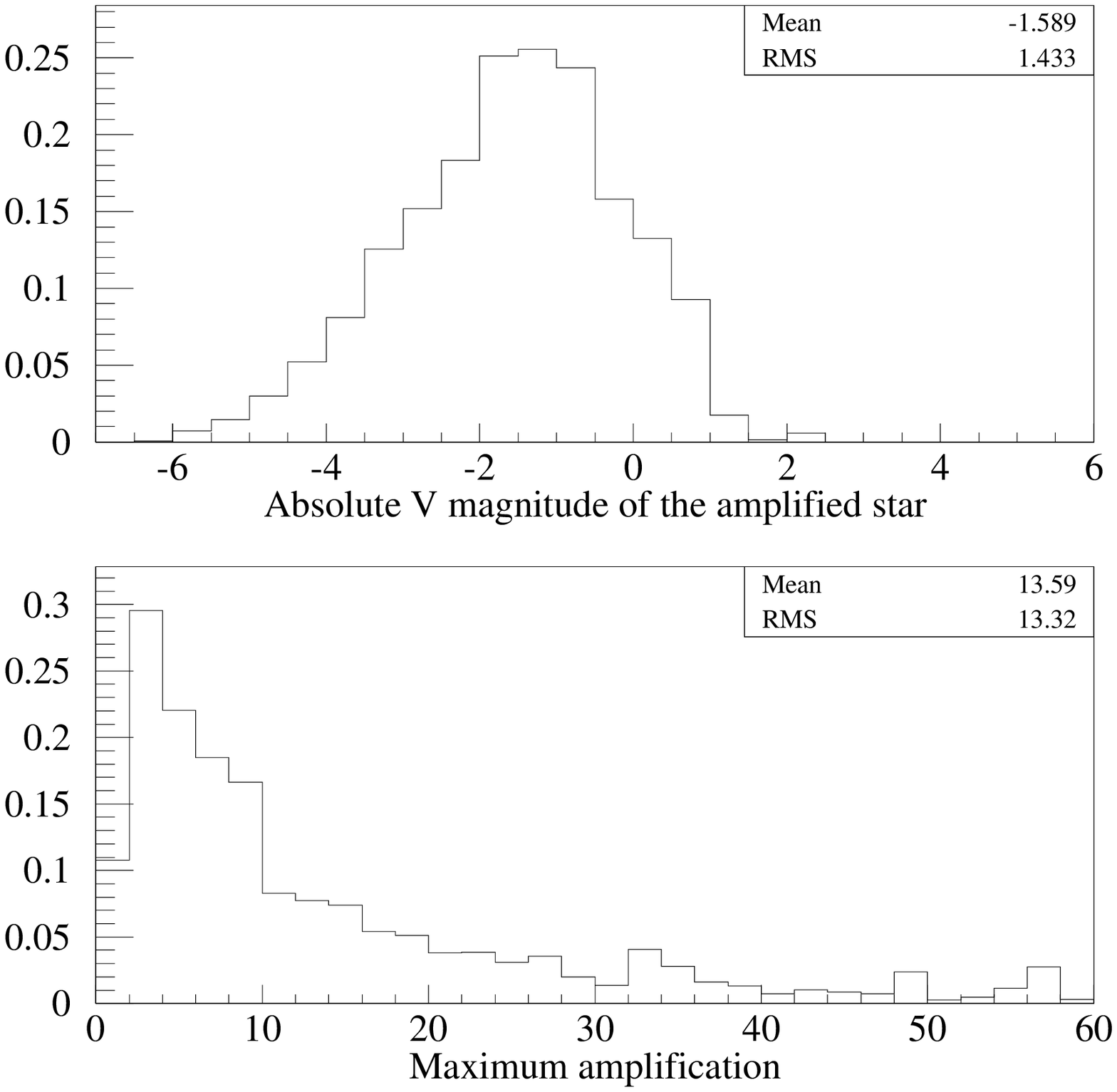,width=.48\textwidth}
\caption[]{{\em Simulated} distributions of the absolute V magnitude%
of the lensed star and of the maximum amplification, for%
``detected'' events.\label{fig02}}%
\end{figure}
}
\newcommand{\figtrois}{
\begin{figure}[ht!]
\centering
\psfig{file=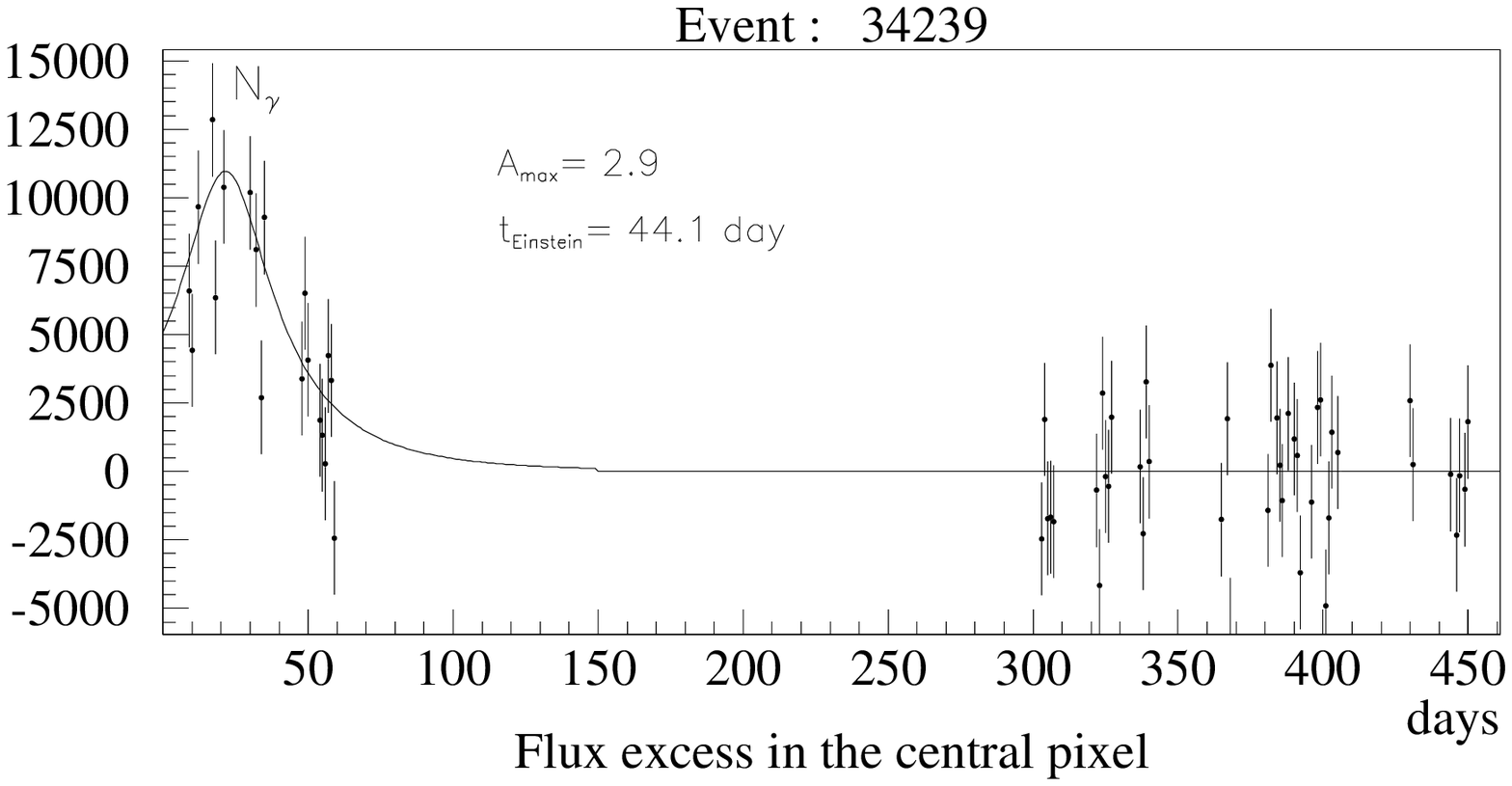,width=.48\textwidth}
\psfig{file=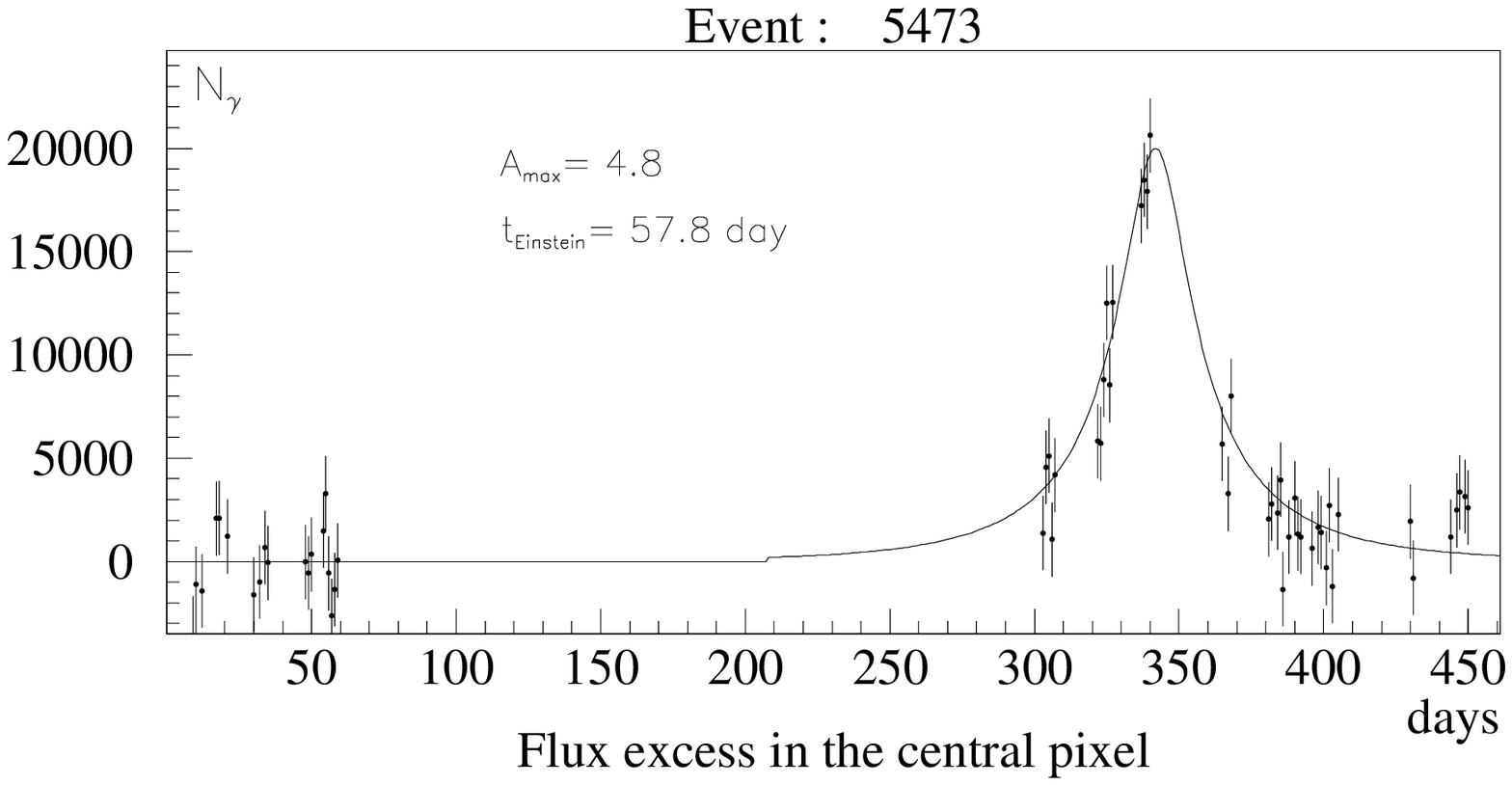,width=.48\textwidth}
\psfig{file=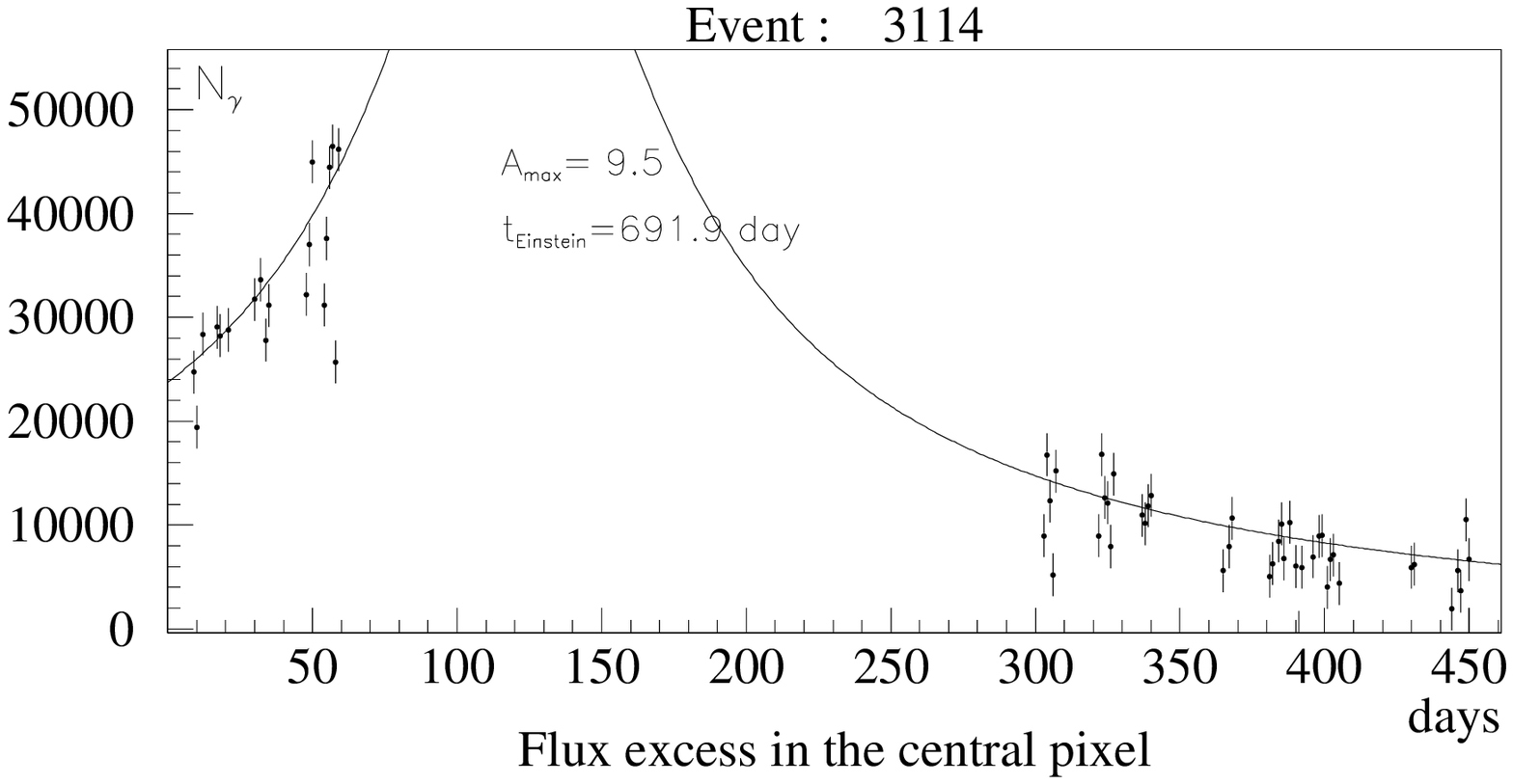,width=.48\textwidth}
\caption[]{Light curves of {\em simulated} ``detected'' microlensing
    events. The solid lines are the theoretical Paczy\'nski curves.\label{fig03}} 
\end{figure}
}
\newcommand{\figquatre}{
\begin{figure}[ht!]
\centering
\psfig{file=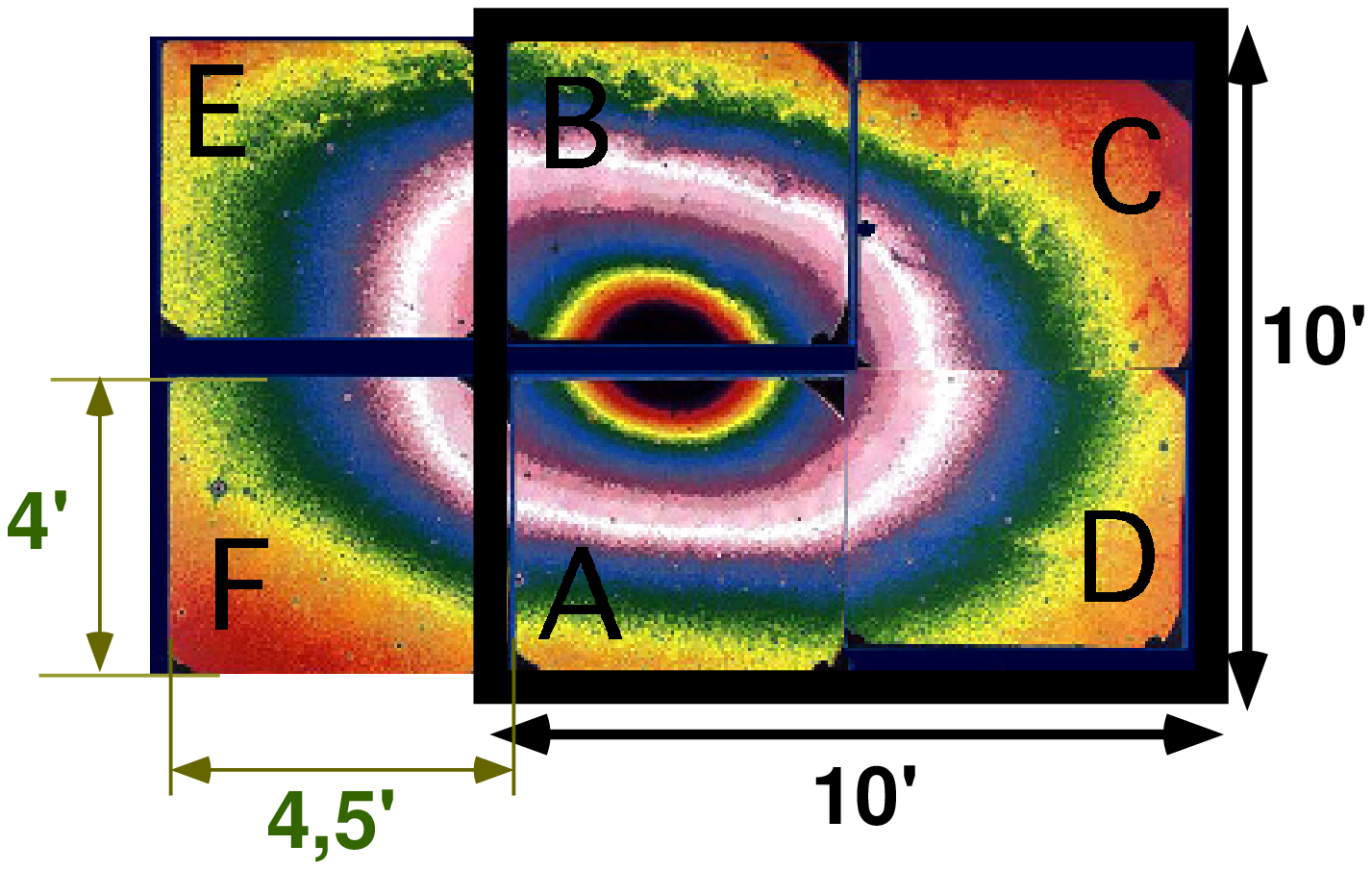,width=.48\textwidth}
\caption[]{Approximate position of fields A to F with respect to M31.\label{fig04}}
\end{figure}
}
\newcommand{\figcinq}{
\begin{figure}[ht!]
\centering
\psfig{file=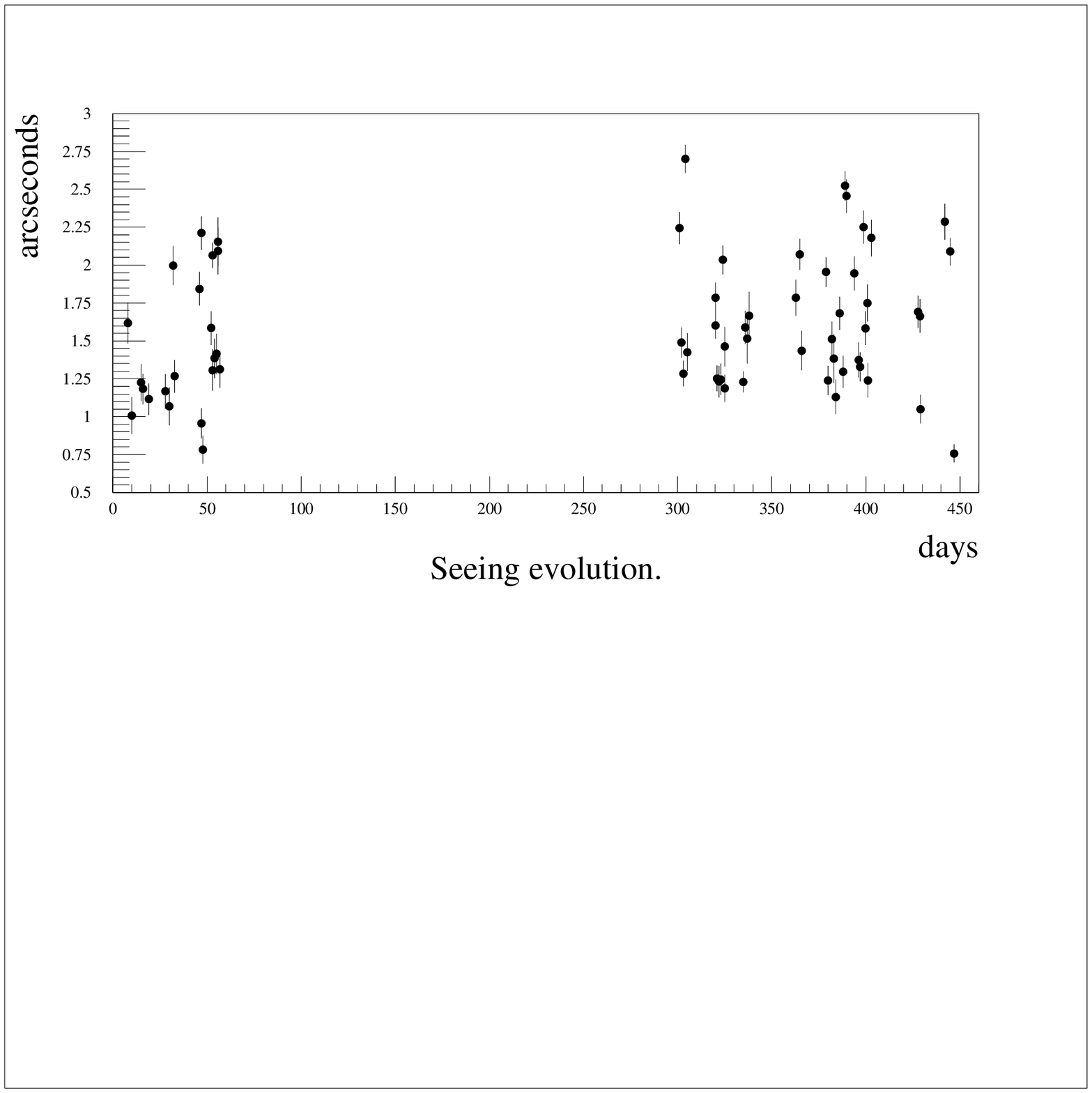,clip=,width=.48\textwidth}
\psfig{file=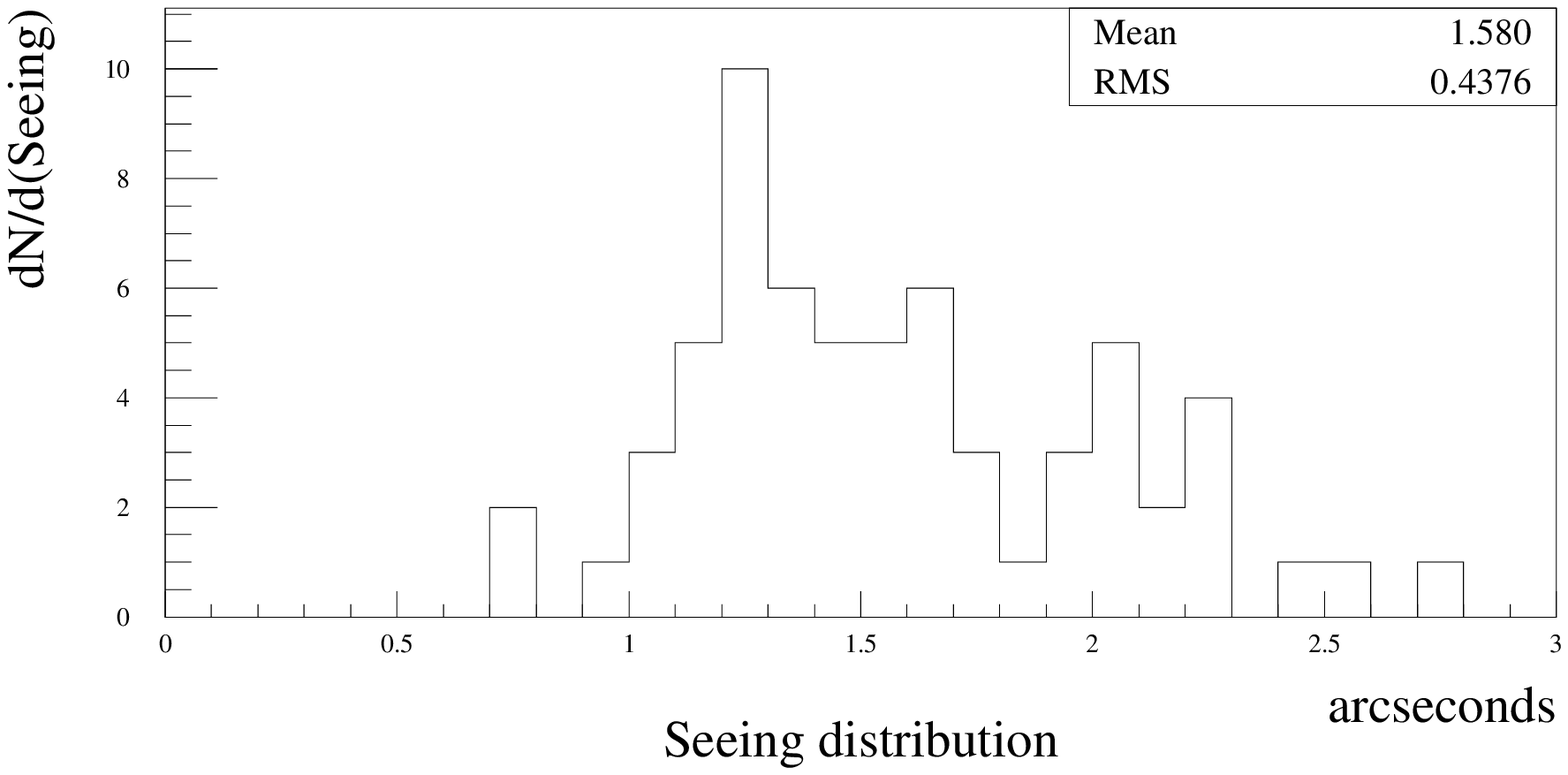,width=.48\textwidth}
\caption[]{The seeing for the 1994 and 1995 runs.\label{fig05}}
\end{figure}
}
\newcommand{\figsix}{
\begin{figure}[ht!]
    \centering
    \psfig{file=fig06.ps,width=.48\textwidth}
\caption[]{Dispersion of the difference of star positions between two images
  after geometric alignment \label{fig06} }
\end{figure}
}
\newcommand{\figsept}{
\begin{figure}[ht!]
\psfig{file=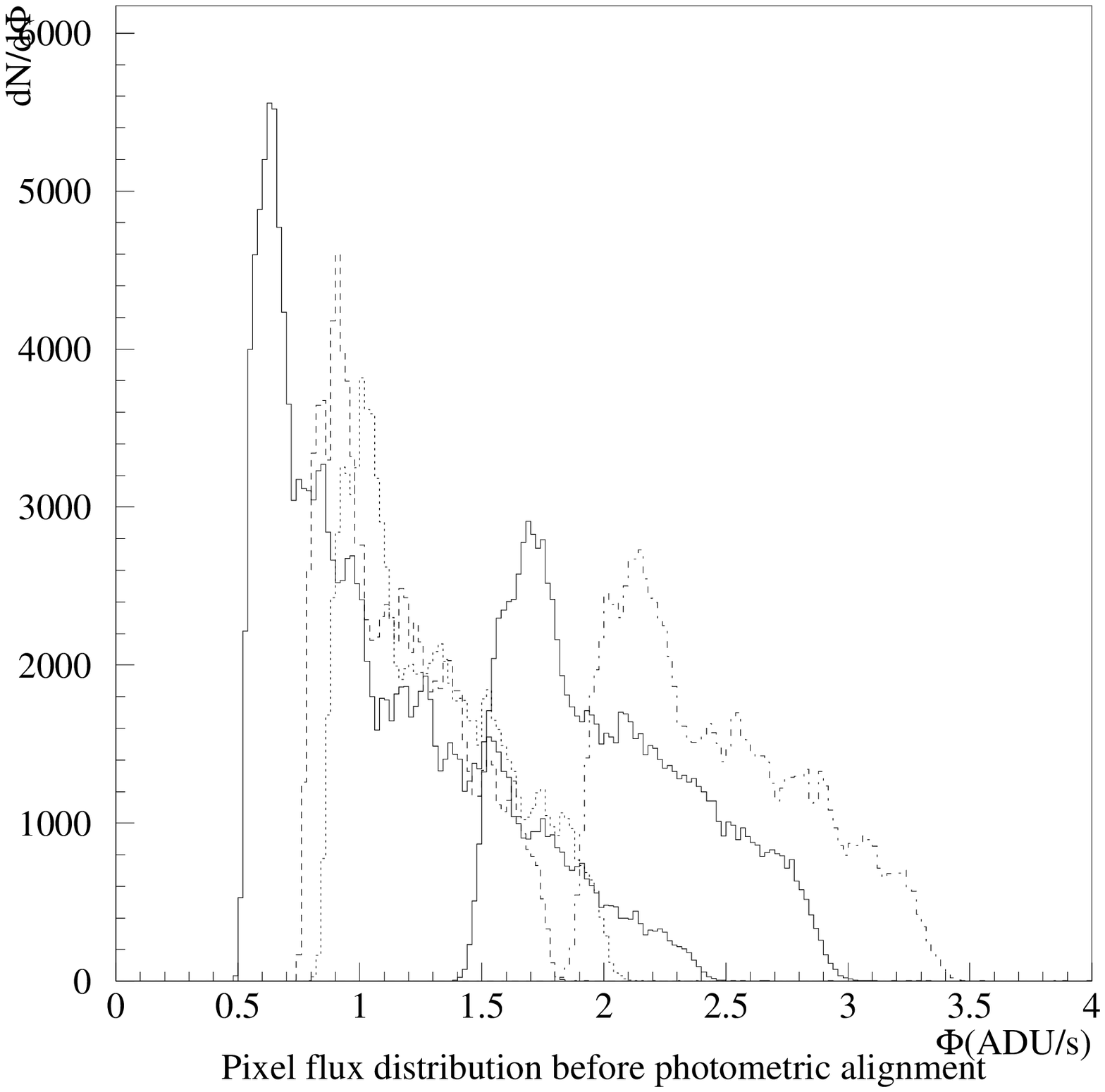,width=.48\textwidth}
\begin{center} a \end{center}
\psfig{file=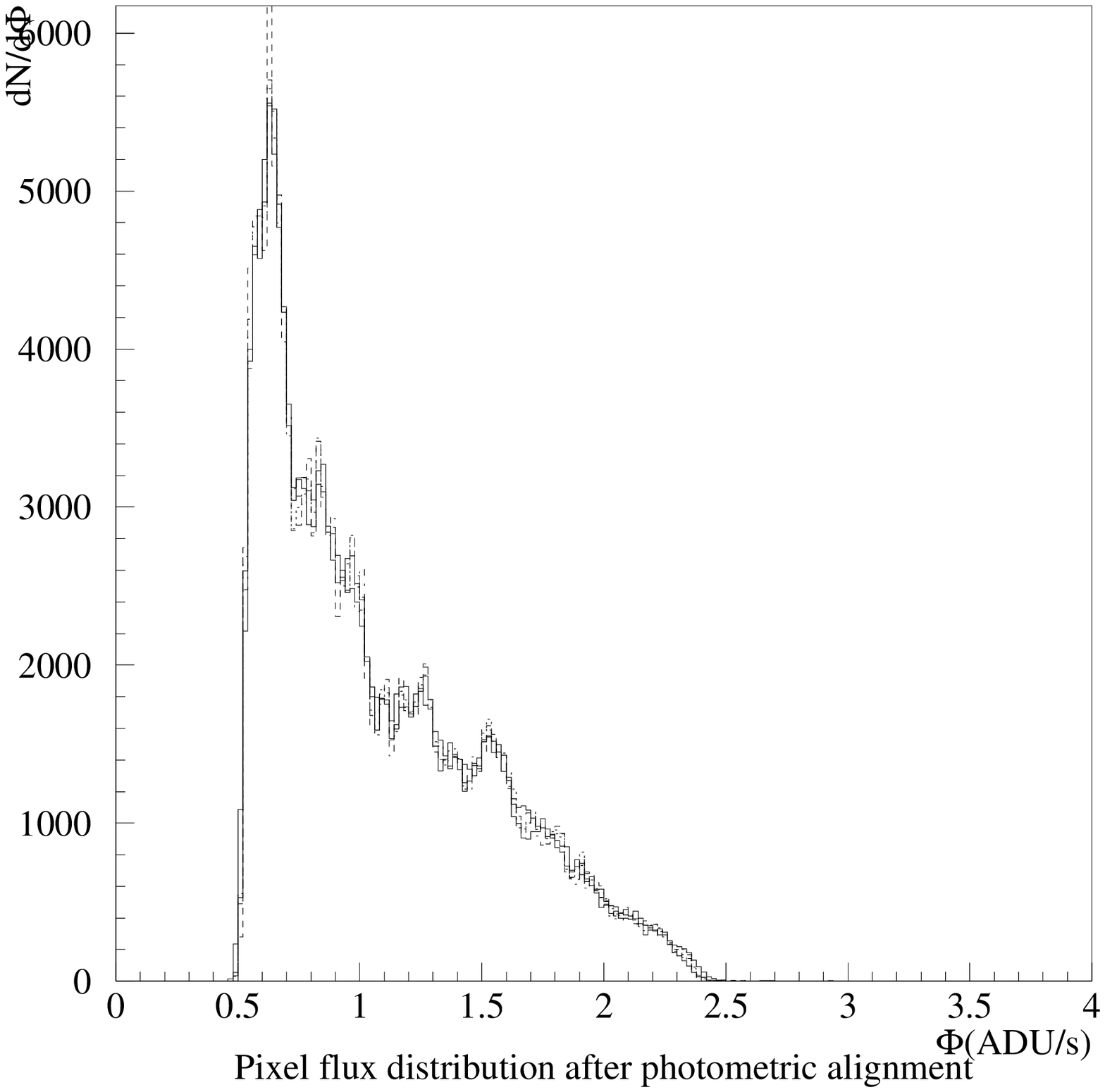,width=.48\textwidth}
\begin{center} b \end{center}
\caption[]{The matching of pixel histograms before (a) 
and after (b) photometric alignment \label{fig07}}
\end{figure}
}
\newcommand{\fighuit}{
\begin{figure}[t!]
\centering
\psfig{file=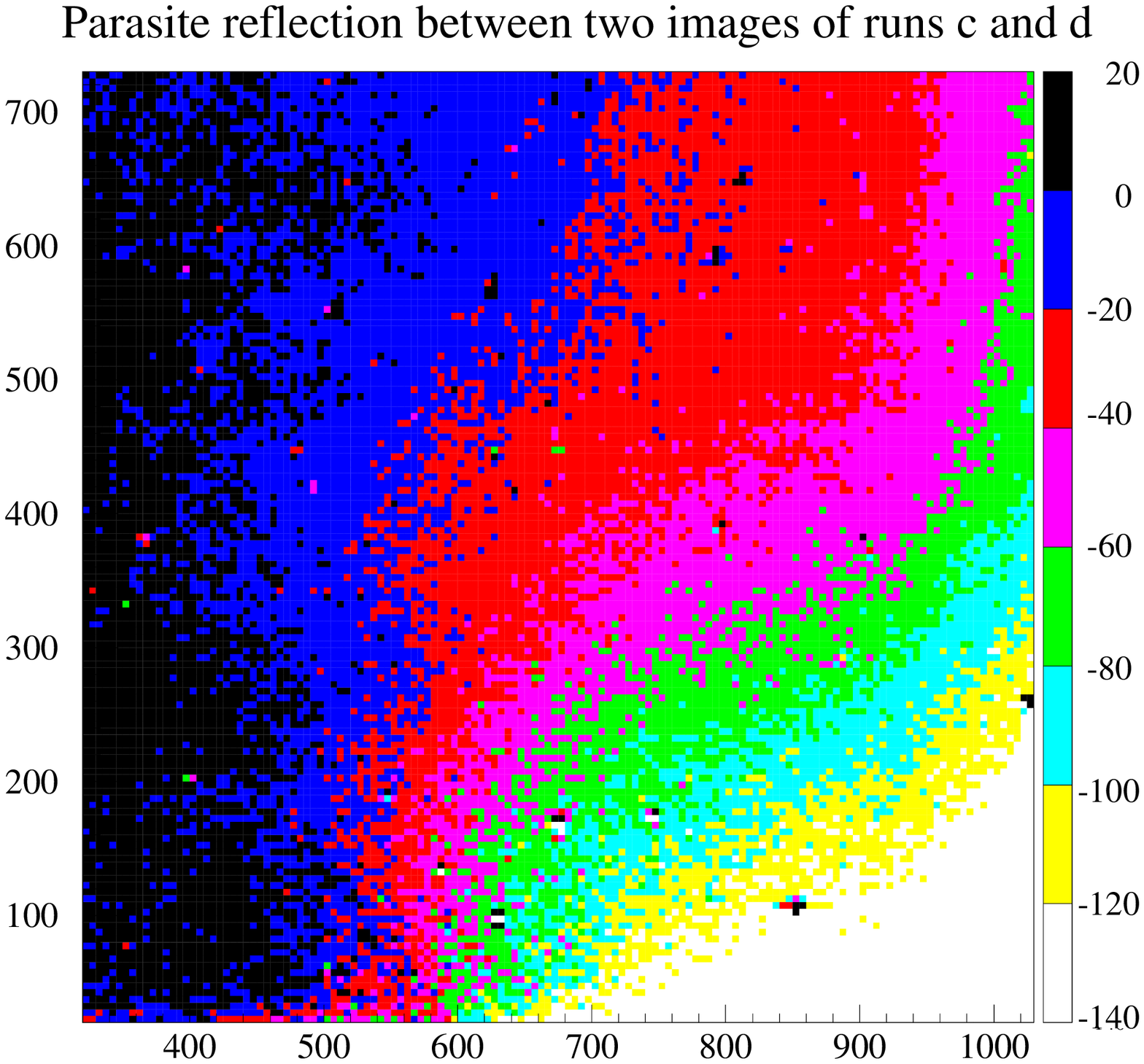,angle=-90,width=0.48\textwidth}
\caption[]{The residual gradient between runs c and d\label{fig08}}
\end{figure}
}
\newcommand{\figneuf}{
\begin{figure}[h!]
  \centering
    \psfig{file=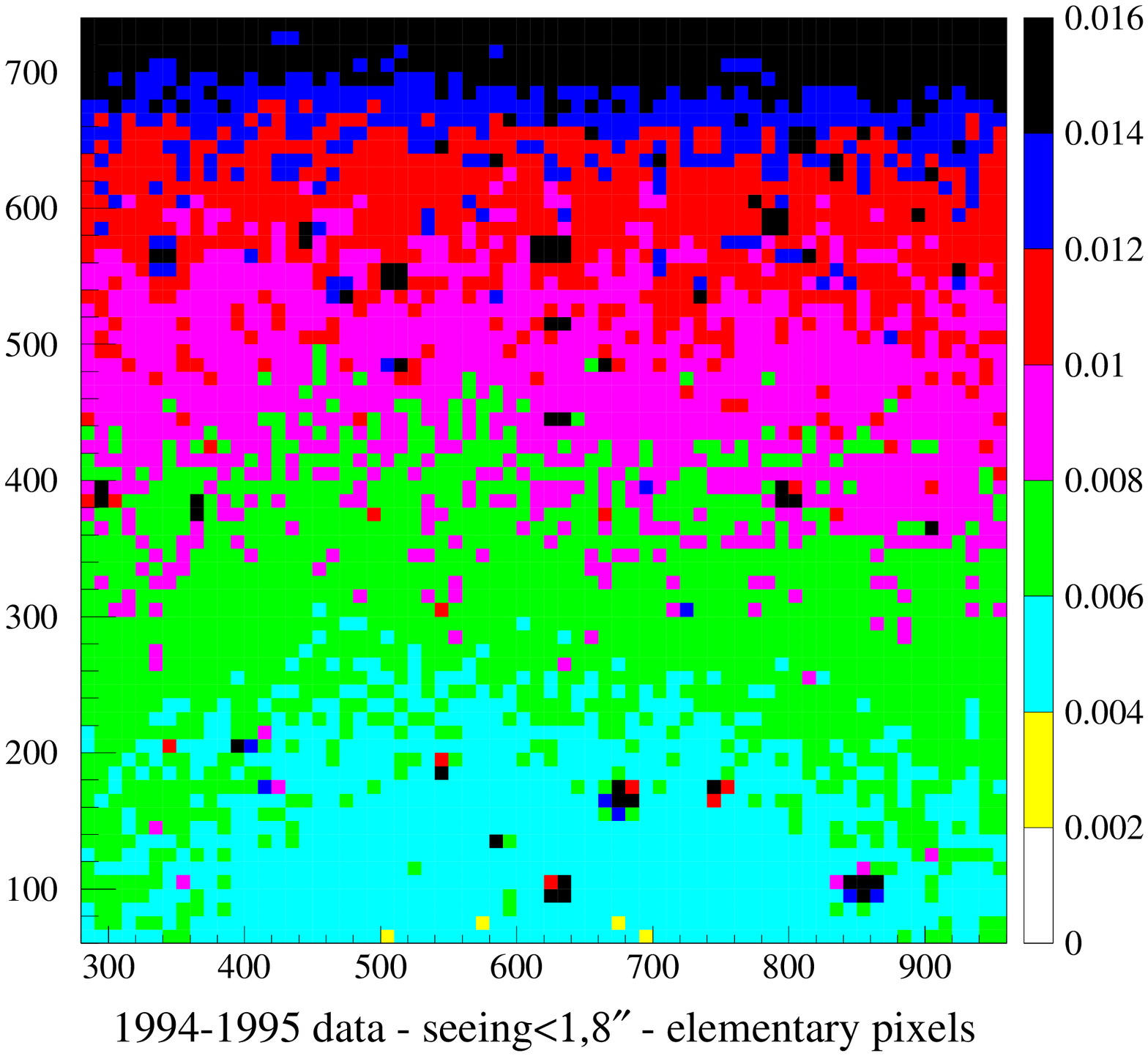,width=.48\textwidth}
    \psfig{file=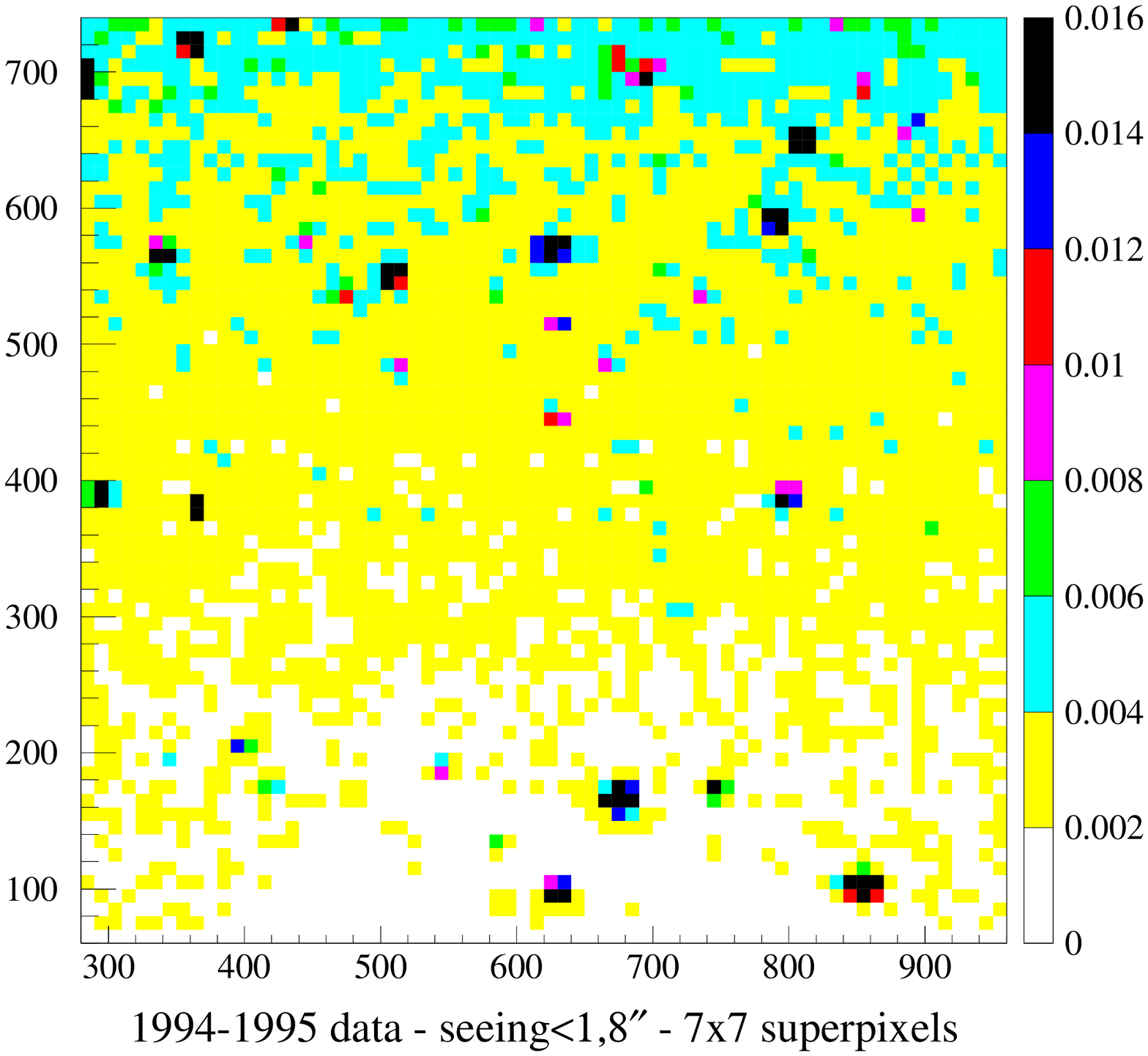,width=.48\textwidth}
  \caption[]{Maps of the relative fluctuation on field A for $(0.3\arcsec)^2$
    elementary pixels (upper map), and $(2.1\arcsec)^2$ super-pixels
    (lower map)\label{fig09}.} 
\end{figure}
}
\newcommand{\figdix}{
\begin{figure}[ht!]
    \psfig{file=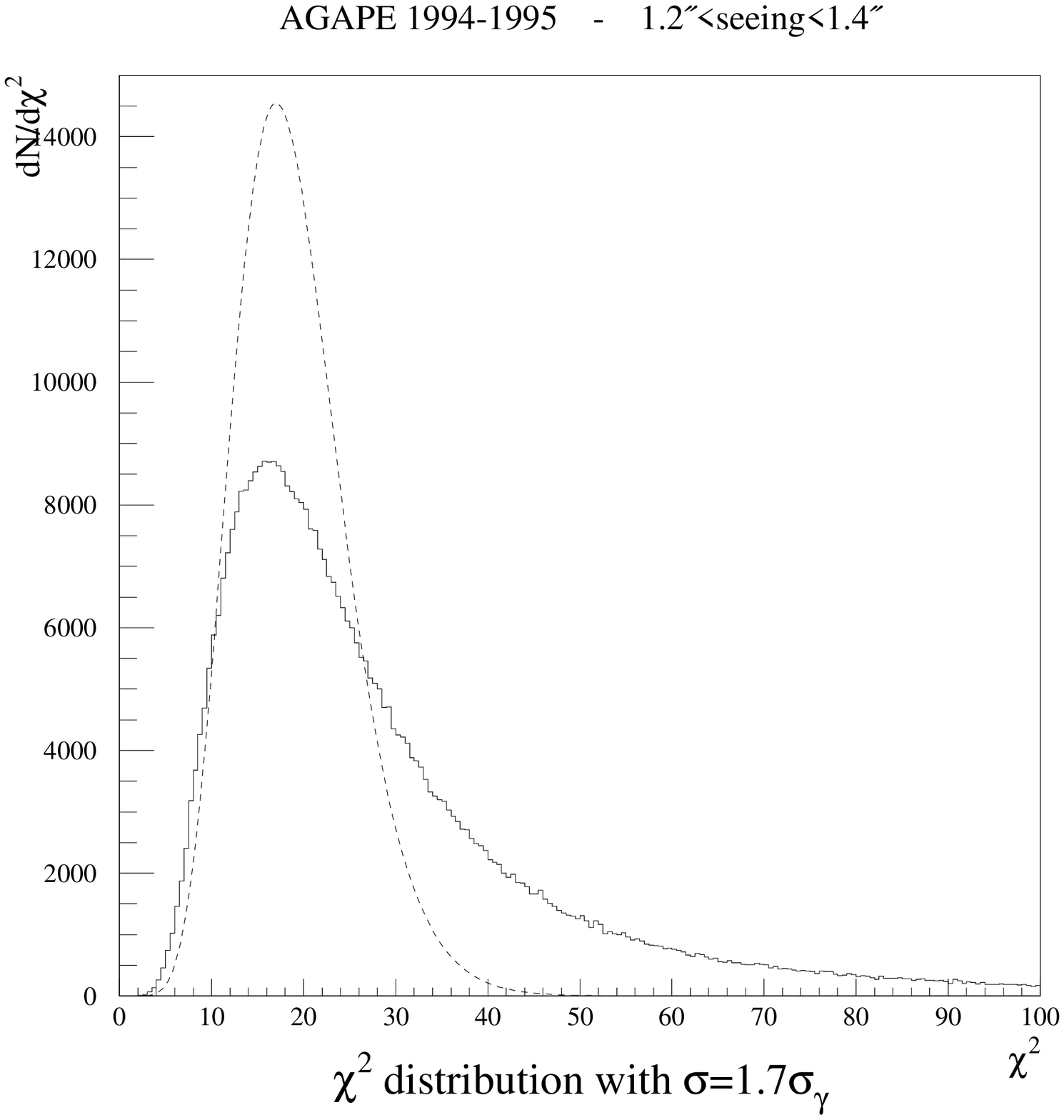,width=.48\textwidth}
   \psfig{file=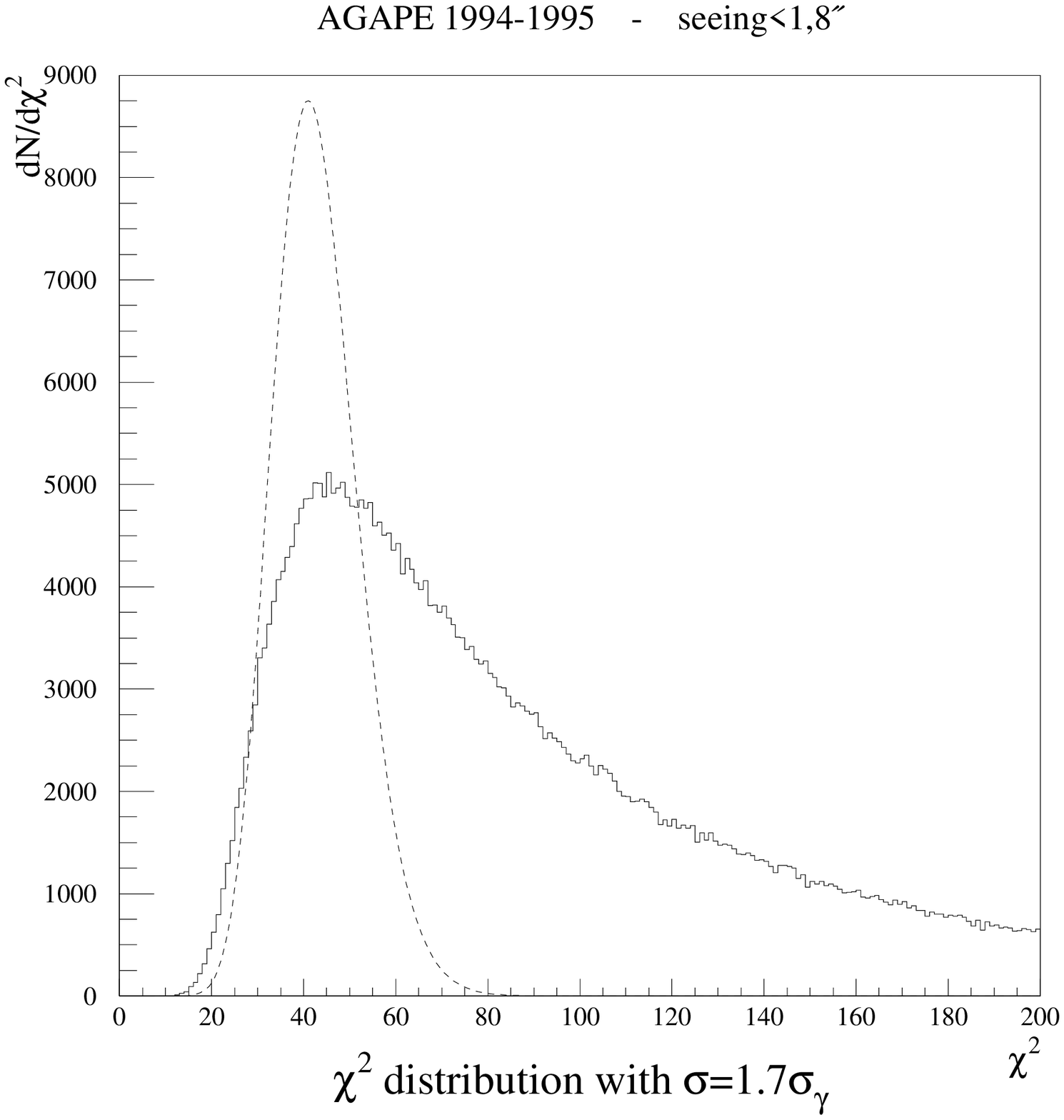,width=.48\textwidth}
  \caption[]{The distribution of $\chi^2$ along light curves of field~A  
    \label{fig10}}
\end{figure}
}
\newcommand{\figonze}{
\begin{figure}[ht!]
  \centering
    \psfig{file=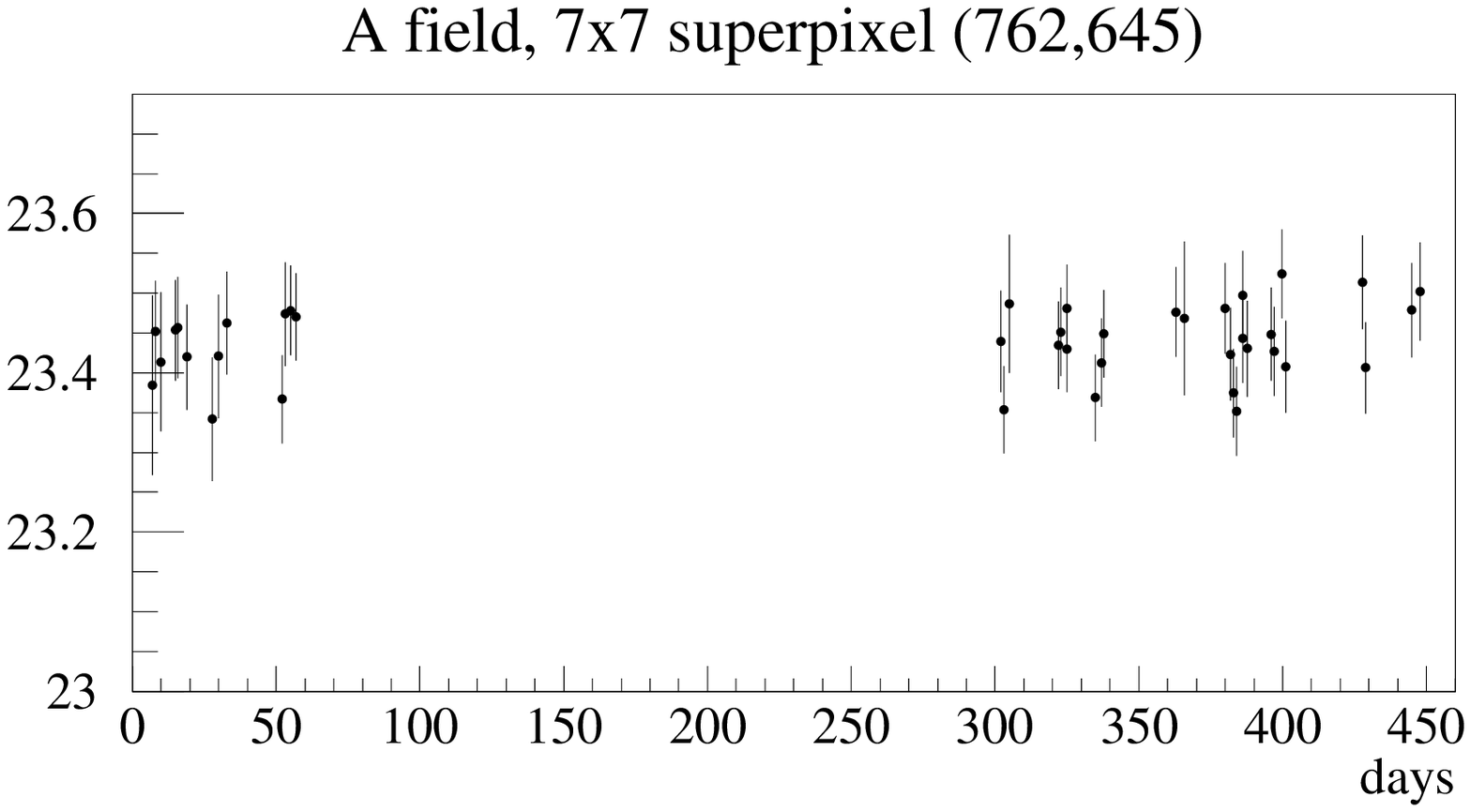,width=0.48\textwidth}
 \caption[]{The light curve of a stable $2.1\arcsec$
    super-pixel. The intensity per super-pixel is in
    ADU/s ($1.1<\mathrm{seeing}<1.8$).\label{fig11}} 
\end{figure}
}
\newcommand{\figdouze}{
\begin{figure}[ht!]
  \centering
    \psfig{file=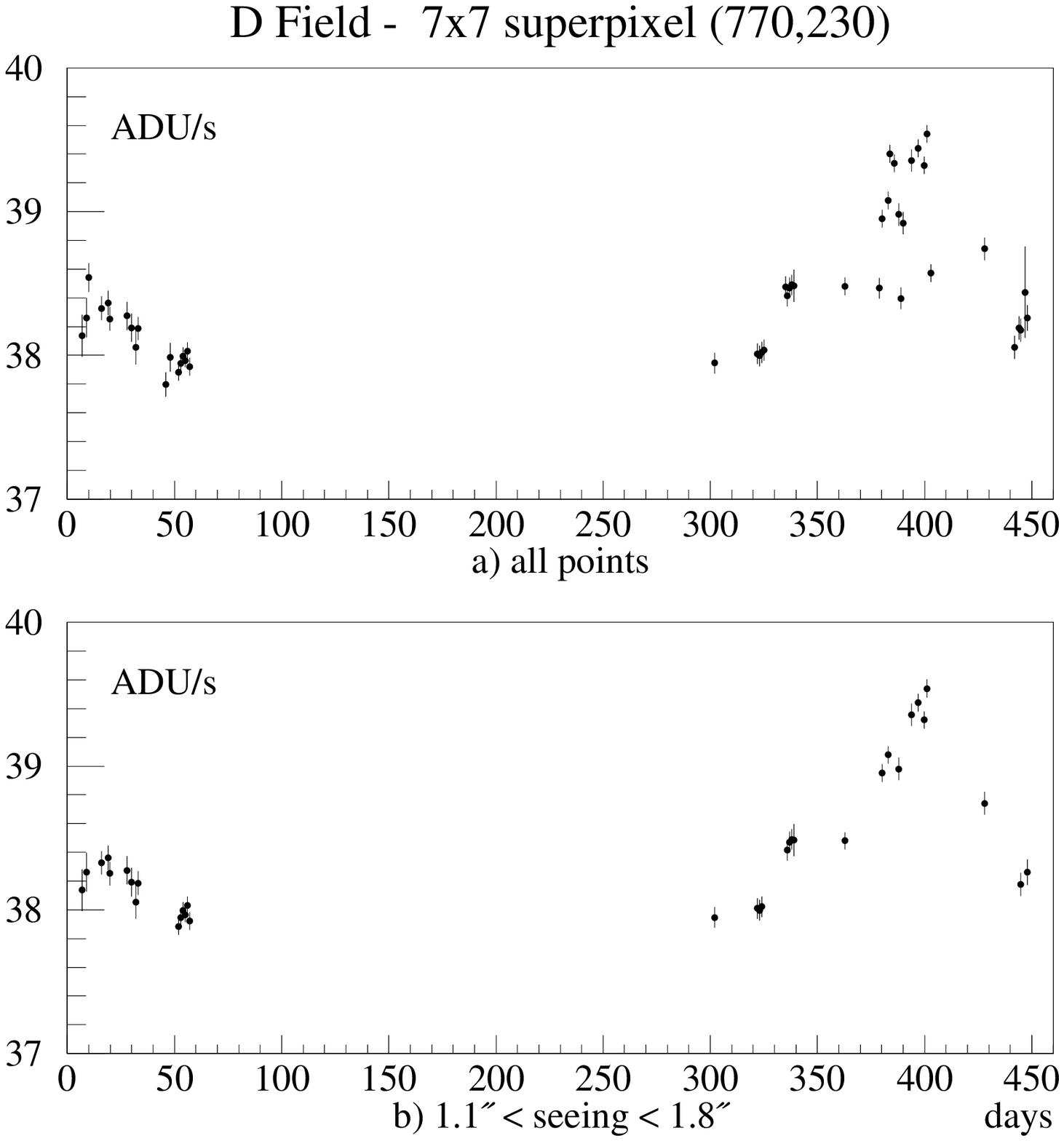,width=0.48\textwidth}
 \caption[]{An example of a variable object.\label{fig12}}
\end{figure}
}
\newcommand{\figtreize}{
\begin{figure}[ht!]
  \centering
    \psfig{file=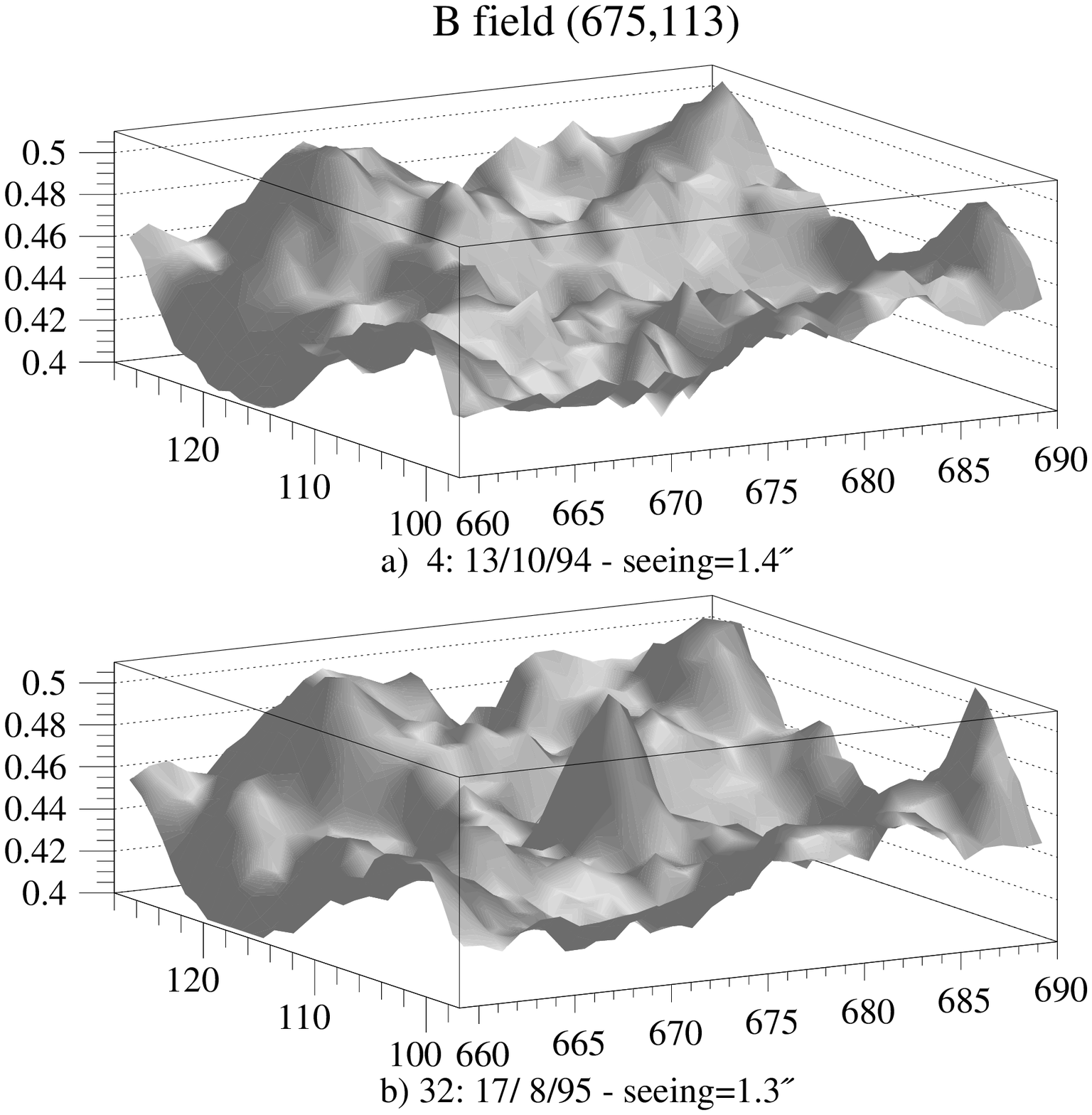,width=0.48\textwidth}
 \caption[]{Appearance of a star. The vertical scale is the intensity
   per elementary pixel, in ADU/s \label{fig13}}  
\end{figure}
}
\newcommand{\figquatorze}{
\begin{figure}[ht!]
  \centering
    \psfig{file=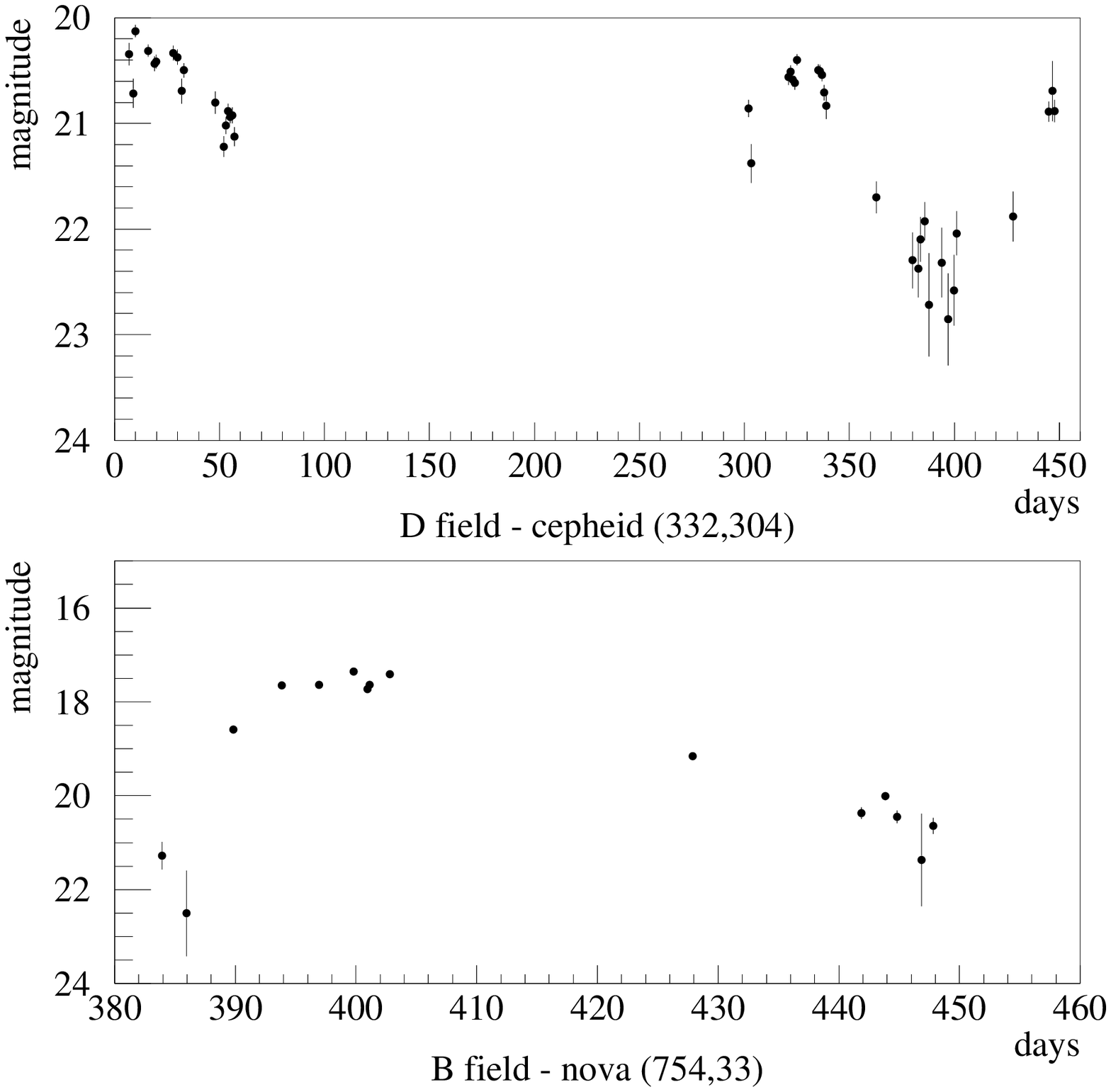,width=0.48\textwidth}
 \caption[]{Likely cepheid and nova.\label{fig14}}  
\end{figure}
}
\newcommand{\figquinze}{
\begin{figure}[ht!]
  \centering
    \psfig{file=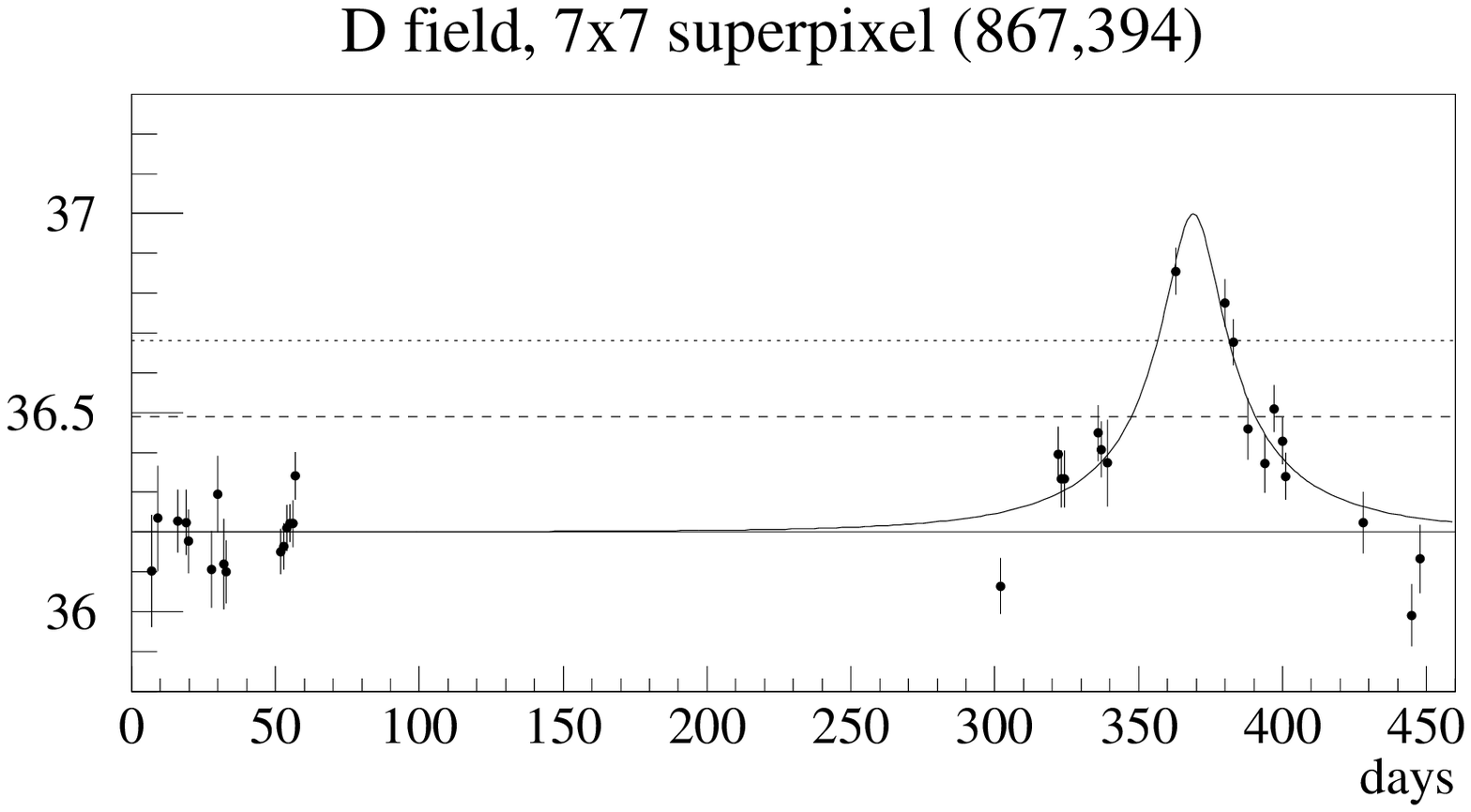,clip=,width=0.48\textwidth}
\caption[]{A possible microlensing event. The solid horizontal line
   is the basis    level of the super-pixel intensity (in ADU/s), the
   dashed line lies    $3\sigma$ above    and the dotted line
   $5\sigma$ above. \label{fig15}}   
\end{figure}
}
\newcounter{saveeqn}%
\newcommand{\alpheqn}{\setcounter{saveeqn}{\value{equation}}%
\stepcounter{saveeqn}\setcounter{equation}{0}%
\renewcommand{\theequation}{\mbox{\arabic{saveeqn}-\alph{equation}}}}%
\newcommand{\reseteqn}{\setcounter{equation}{\value{saveeqn}}%
\renewcommand{\theequation}{\arabic{equation}}}%

\begin{document}
\thesaurus{06(10.08.01;12.03.3;12.04.1;12.07.1)}
\title{
AGAPE: a search for dark matter towards M~31 by
microlensing effects on  unresolved stars
\thanks{Based on data collected with the 2m Telescope Bernard
Lyot (TBL) operated by INSU-CNRS and Pic du Midi Observatory (USR
5026).\protect\newline
The experiment was funded by IN2P3 and INSU of CNRS}}

\author{R. Ansari\inst{1} \and M. Auri{\`e}re \inst{2} \and P. Baillon \inst{3}
\and A. Bouquet \inst{4, 5} \and G. Coupinot \inst{2} \and Ch. Coutures \inst{6}
\and C. Ghesqui{\`e}re \inst{5} \and \\ Y. Giraud-H{\'e}raud \inst{5} \and
P. Gondolo\thanks{Supportted in part by the European Community (EC
  contract no. CHRX-CT93-0120)}
\inst{4, 7} \and J. Hecquet \inst{8} \and J. Kaplan \inst{4, 5} \and
Y. Le Du \inst{5} \and A.L. Melchior\thanks{Supported in part by Fondation
  Singer Polignac}
\inst{5} \and \\ M. Moniez\inst{1} \and J.P. Picat \inst{8} \and G. Soucail \inst{8}}

\offprints{J. Kaplan, kaplan@cdf.in2p3.fr}

\institute{Laboratoire de l'Acc{\'e}l{\'e}rateur Lin{\'e}aire, Universit{\'e} Paris-Sud,
91405 Orsay, France
\and Observatoire Midi-Pyr{\'e}n{\'e}es, unit{\'e} associ{\'e}e au CNRS (UMR 5572),
62500 Bagn{\`e}res de Bigorre, France 
\and CERN, 1211 Gen{\`e}ve 23, Switzerland
\and Laboratoire de Physique Th{\'e}orique et Hautes Energies, Universit{\'e}s Paris
6 and Paris 7, unit{\'e} associ{\'e}e au CNRS (URA 280), 2, Place Jussieu, 75251
Paris Cedex 05, France
\and Laboratoire de Physique Corpusculaire, Coll{\`e}ge de France, Laboratoire
associ{\'e} au  CNRS-IN2P3 (URA 6441), 11 place Marcelin Berthelot, 75231
Paris Cedex 05, France 
\and SPP/DAPNIA, CEN Saclay, 91191 Gif-sur-Yvette, France
\and Department of Physics, University of Oxford, 1 Keble Road,
Oxford, OX1 3NP, United Kingdom 
\and Observatoire Midi-Pyr{\'e}n{\'e}es, unit{\'e} associ{\'e}e au CNRS (UMR 5572), 14
avenue Belin, 31400 Toulouse, France}

\date{Accepted December 09 96}

\maketitle

\begin{abstract} 
M~31 is a very tempting target for a microlensing search of compact objects in
galactic haloes. It is the
nearest large galaxy, it probably has its own dark halo, and its
tilted position with respect to the line of sight provides an
unmistakable signature of microlensing. However most stars of M~31 are
not resolved and one has to use the ``pixel method'': monitor the
pixels of the image 
rather than the stars. AGAPE is the implementation of this
idea. Data have been collected and
treated during two autumns of observation at the 2 metre telescope of Pic
du Midi. The process of geometric and photometric alignment, which must
be performed before constructing pixel light curves, is
described. Seeing variations are minimised by working with
large super-pixels ($2.1 \arcsec$) compared with the average seeing. A
high level of stability of pixel fluxes, crucial to the approach, is
reached. Fluctuations of super-pixels do not exceed 1.7 times the
photon noise which is 0.1\% of the intensity for the brightest
ones. With such stable data, 10 microlensing events are expected for a
full ``standard halo''. With a larger field, a regular and short time
sampling and a long lever arm in time, the pixel method will be a very
efficient tool to explore the halo of M~31.
\end{abstract}

\section{The background of the AGAPE search}

\subsection{Dark matter in galaxies}

The presence of a large amount of unseen matter is a very old astrophysical
problem (Oort \cite{OORT}, Zwicky \cite{zwicky}) but its importance
was widely recognised only in the seventies (Ostriker, Peebles \&
Yahil \cite{OPY}, Faber \& Gallagher 
\cite{FG}). Actually there are several ``dark matter problems''
on different scales: stellar systems, individual galaxies, clusters
and superclusters of galaxies, up to cosmological scales. Dark matter
appears also necessary to understand large structures formation. For a recent
review on these subjects, see Dolgov~(\cite{dolgov}). Many
observations suggest that spiral galaxies are embedded in  
massive dark haloes (Kormandy \& Knapp \cite{KK}, Trimble
\cite{trimble}). The most conspicuous evidence for such haloes is the
rotation curve of galactic disks, which does not decrease near the outskirts
of galaxies. If the mass density and surface brightness profiles were
similar, the rotation curve should fall according to Kepler's law. 

From the rotation curves of spiral galaxies, one can estimate the amount of dark 
matter within 2 Holmberg radii to be larger by one order of magnitude
than the amount of luminous matter, but the
shape of dark haloes is unknown. Several lines of argument
point towards a more or  
less spherical distribution, such as the existence of galaxies with a
rapidly rotating polar ring, the stability of the disk of spiral
galaxies against bar formation (Ostriker \& Peebles \cite{OP}) or the
distribution of the globular clusters (Harris \& Racine~\cite{HR}). The sphere
seems often flattened in the direction of the rotation axis (for a
recent review, see Sackett \cite{SACKETT} and references therein).

The nature of dark haloes remains also unknown.
Many candidates have been proposed, either baryonic 
or not, ranging from light neutrinos to very heavy black holes of 10$^6 \;
M_\odot$, but it is out of the scope of this paper to review them
extensively (for recent reviews, see for instance Dolgov~\cite{dolgov} and Griest~\cite{griest2}) 
. Nevertheless,
we can mention some unconventional views, such as the modified 
Newton dynamics (Bekenstein \& Milgrom \cite{MOND}), or cold molecular
hydrogen as the constituent of dark haloes of 
spiral galaxies (Pfenniger, Combes \& Martinet \cite{PCM}).

\subsection{Baryonic dark matter}

Although the subject of primordial abundances has recently
become rather confused, Big Bang Nucleosynthesis indicates that the density of
baryonic matter in the universe is probably around 10 times larger
than that
seen as stars or interstellar gas (for a recent discussion, see for
instance Cardall \& Fuller \cite{BBNS} and references therein). But 
the Cosmological Standard Model gives no hint as to the 
location of this baryonic matter and its relative distribution between
galactic haloes and intergalactic medium in clusters of
galaxies. 

It has been suggested that galactic dark matter could be essentially 
made of compact baryonic objects such as low mass 
stars or brown dwarfs.
Brown dwarfs are stars too light ($M < 0.08 M_\odot$) for the
gravitational pressure to fire nuclear reactions and are a natural
candidate for the constituent of galactic haloes (Carr, Bond \& Arnett
\cite{CBA}). It is considered that they should be heavier than
10$^{-7} M_\odot$ lest they would  evaporate too quickly (De R{\'u}jula, Jetzer \&
Mass{\'o} \cite{EVAP}). Such objects should 
most easily be seen in the red and infrared bands (Kerins \& Carr \cite{KC}).
A few may have been in fact observed, some orbiting brighter compagnons:
GD 165B (Zuckerman \& Becklin \cite{ZB}) and Gl229B (Nakajima et
al. \cite{NAKA}, Allard et al. \cite{FALLARD}), as well as others free flying 
in the Pleiades cluster: PPl 15 (Stauffer, Hamilton \& Probst
\cite{SHP}), Teide 1 (Rebolo, Zapaterio Osorio \& Mart{\'\i}n
\cite{RZM}) and Calar 3 (Zapaterio Osorio, Rebolo \& Mart{\'\i}n
\cite{ZRM}). Both PPl 15 and Teide~1 have residual Lithium, and Calar 3
resembles Teide 1 like a twin. (Basri, Marcy \& Graham \cite{BMG},
Mart{\'\i}n, Rebolo \& Zapaterio Osorio \cite{MRZ}). 

\subsection{Gravitational microlensing}

Direct searches for brown dwarfs can at best explore the solar
neighbourhood. To detect them further out, it was proposed a
few years ago by Paczy\'nski (\cite{pacz}) to search for dark objects through
gravitational lensing. When a compact object passes near the line of
sight of a background star, the luminosity of this star will be
temporarily increased in a characteristic way. 

Several experiments have been implementing this idea since 1990 and
have indeed seen microlensing events. Two groups have been looking towards
the Magellanic Clouds: the EROS collaboration (Aubourg et
al. \cite{EROSa}, Ansari et al. \cite{EROSb}, Milsztajn \cite{MILS}) and the
MACHO collaboration (Alcock et al. \cite{MACHOa,MACHOb}, Bennett
\cite{BENN}). Microlensings have also been
searched for in the direction of the galactic bulge by three groups: OGLE
(Udalski et al. \cite{OGLEa, OGLEb}), MACHO (Alcock et
al. \cite{MACHOBULGE}, Sutherland \cite{SUTH}) and DUO (Alard, Mao \&
Guibert \cite{DUO}, Alard \cite{ALARD}), who have
observed a large number of events. The microlensing phenomenon can now be
considered as established.

However, the number of events towards the Large Magellanic Cloud (LMC) is lower than expected,
50\% or less of what one would expect with a standard spherical
halo (Bennett \cite{BENN},  Milsztajn \cite{MILS}), but
statistics remain very poor. Moreover, with only one line of sight,
it is very difficult to disentangle the various parameters which enter
in a galactic halo model: density, velocity distribution,
mass distribution, flattening. 

MACHO will continue for two more years
and the upgrade of EROS (Couchot \cite{COUCH}) will start operation
soon. However, the ``classical'' technique used in these
experiments does not allow to explore other directions through the halo,
because the two Magellanic Clouds are the only possible targets with enough
resolved stars.

\subsection{Going further, the ``pixel method''}
It is thus tempting to look at rich fields of stars further out,
such as the M~31 galaxy. But most stars of M~31 are not resolved and a new
technique must be developed. Such a technique, the ``pixel method'',
has been proposed and implemented by us
(Baillon et al. \cite{BBGK1,BBGK2}, Ansari et al. \cite{ROME}).
A similar idea, relying on image subtraction, has been independantly
proposed by (Crotts \cite{crotts}), and implemented by the
Columbia-VATT collaboration (Tomaney \& Crotts \cite{TAC},
Tomaney \cite{tomaney}).

The method we propose is the following: in a dense field of stars,
many of them contribute to each pixel. However if one {\em unresolved}
star is sufficiently magnified, the increase of the total flux of the pixel
will be large enough to be detected. Therefore, instead of monitoring
individual stars, we propose to follow the luminous intensity of the pixels of
the image. Then {\em all} stars in the field, and not the only few
resolved ones, are candidates for a micro-lensing, so that the
event rate is potentially much larger. Of course, only the brightest
stars will be amplified enough to become detectable above the
fluctuations of the background, unless the amplification is very high
and this occurs very seldom. In a galaxy like M~31, however, this is
compensated for by the very high density of stars, 
and indeed various
evaluations (Baillon et al. \cite{BBGK2}, Jetzer \cite{jetzer94},
Colley \cite{colley}, Han \& Gould \cite{HG}) 
show that a fair number of events should be detectable.

This paper is devoted to the description of AGAPE (Andromeda
Gravitational Amplification Pixel Experiment), which implements this
idea in the direction of M~31, on data taken in autumns 1994 and 1995 at
the 2 metre telescope Bernard Lyot (TBL) at Pic du Midi Observatory in
the French Pyr{\'e}n{\'e}es. 

In section \ref{method}, after recalling the principles of the method,
(introduced in Baillon et al. \cite{BBGK1} and \cite{BBGK2}), we give analytic
evaluations of the number of events expected. Although these analytic
estimates can at best be very rough, they provide useful qualitative
insights. To get reliable estimates in the true observational
conditions, we resort to Monte-Carlo simulations. 

In section~\ref{observ} we describe the telescope,  the detector, the
conditions and the course of the observations. Section~\ref{treatment}
is devoted to the 
geometric and photometric alignments of successive images and to the
absolute photometry. 
In section~\ref{results} we show that the high level of stability
reached on the average super-pixel (a group of $7\times7$ elementary
pixels) allows us to detect variable objects that would have been very
difficult to see otherwise. The detailed analysis of the variations we
detect will be the subject of separate publications. 

The pixel method should also give interesting results in the bar of the LMC, and we have
started to analyse the data of the EROS collaboration in this framework
(Melchior \cite{THAL}). The results will also be published elsewhere.

\section{The pixel method \label{method}}

The photon
flux of an  individual
star, $F_{\mathrm{ star}}$,  is spread among all pixels of the seeing
disk and only part of this light, the seeing fraction $f$,  reaches
the pixel nearest to the centre of the star:
\beq
F_{\mathrm{star,\; pixel}} = f \times \ F_{\mathrm{ star}}.
\eeq
In a crowded field such as M~31, the light flux 
$F_{\mathrm{pixel}}$ on a pixel comes from  the 
many stars in and around it, plus the sky background.
\beq
F_{\mathrm{pixel}} = F_{\mathrm{neighbouring\ stars}} + F_{\mathrm{sky background}}  
\eeq
  If the luminosity of a particular star is amplified by a factor $A$, the pixel flux
increases by:
\beq
\Delta F_{\mathrm{pixel}} = (A-1) \ f \ F_{\mathrm{star}}\,.
\eeq
The amplification of the star
luminosity can be detected if 
the flux on the pixel nearest to its centre rises sufficiently high
above the rms fluctuation $\sigma_{\mathrm{ pixel}}$:
\beq
\Delta F_{\mathrm{pixel}} > Q\ \sigma_{\mathrm{pixel}}\,. \label{detect}
\eeq
Of course, to be detected, a lensing event should be visible on several
exposures. One therefore typically requires that condition (\ref{detect}) be
verified for at least 3 consecutive pictures with $Q = 3$ and with $Q = 5$ for
at least one of the three.

Seeing variations induce unwanted fluctuations of the pixel fluxes. To
minimise this problem, and to collect most of the light of any varying
object, 
we replace {\em each} elementary pixel by a ``super-pixel'' centered on
it. Each  super-pixel is a square of $n\times n$ elementary pixels. The
size of the square is chosen large
enough to cover the whole seeing disk in most cases, but also not too
large, to avoid dilution of a variable signal when it occurs. We have
also tried to replace each pixel by an average of the neighbouring
pixels weighted with the point spread function (PSF), as it is known to
maximize the signal to noise ratio at the center of a star on a given
image. However, for this very reason, it turns out that this procedure  
amplifies considerably the fluctuations in time due to seeing variations and
therefore it is not appropriate for our method.

\subsection{Microlensing tests}
All of the classical tests can be applied to discriminate microlensing events against
other sources of light variations.
\paragraph{Uniqueness} The probability of a microlensing occurring twice on 
stars contributing to the same pixel is very weak, and it is safe to reject all non
unique events.
\paragraph{Symmetry} Except in the case of a multiple lens or star, the light
curve should be symmetric in time around the maximum amplification.
\paragraph{Achromaticity} Gravitational lensing is an achromatic
phenomenon. However, the lensed star has not, in general, the same colour as the
background and only the luminosity increase is achromatic (assuming
constant seeing):
\beq
\frac{\Delta F_{\mathrm{pixel}}^{\,\mathrm{red}}}{\Delta
  F_{\mathrm{pixel}}^{\mathrm{\,blue}}} =
\frac{F_{\mathrm{star}}^{\,\mathrm{red}}}{F_{\mathrm{star}}^{\,\mathrm{blue}}}
= \mbox{constant in time}
\eeq
\paragraph{A specific signature: forward-backward asymmetry} It has
been pointed out by Crotts (1992) that M~31 provides a unique test of
microlensing. As this galaxy is tilted with respect to our line of
sight, the rate of microlensing should be higher for those regions of
its disk which are on the far side, because they lie behind a
larger fraction of the halo of M~31 and should undergo microlensing
more often. 
Therefore, one expects a forward-backward asymmetry in
the distribution of microlensing events, which cannot be faked by intrinsically
variable objects.

 \subsection{Expected number of events \label{NOV}}

Most~basic formulae can be found in
Griest (\cite{griest}) and De R{\'u}jula, Jetzer \& Mass{\'o} (\cite{DJM2}). We only recall those few that we shall explicitly
need.  

The amplification $A$ is related to the distance of the lens to the line of
sight $u\,R_E$ ($R_E$ is the Einstein radius) by the relation:
\beq
A=\frac{u^2+2}{u\sqrt{u^2+4}}\label{eq:udea}
\eeq
We detect the variation with the time $t$ of this amplification when a
lens passes near the line of sight with a transverse velocity
$v_\perp$. Then 
\beq
u(t) = \sqrt{\left(\frac{t-t_0}{t_E}\right)^2 +u_0^2},
\eeq 
where $t_0$ and $u_0$ are the time and distance of maximum
amplification, and the Einstein time, $t_E=R_E/v_\perp$, is the time it takes for
the lens to cover one Einstein radius.
 
The rate of events where the amplification is larger than a
definite value $A$ is proportional to the amplification radius $u(A)$
(obtained by inversion 
of equation (\ref{eq:udea}))
\beq
\Gamma = \Gamma_0\, u(A)\, ,
\eeq
where $\Gamma_0$ is the rate of events for which the impact parameter 
gets smaller than the Einstein radius and the amplification exceeds
1.34. Note that the rate $\Gamma$ is linear in the amplification radius
$u(A)$, because it counts the number of stars that enter the area
inside $u(A)$ per unit of time.

\paragraph{
Lenses in the Milky Way halo} The simple evaluations that
follow can only be made for lenses in the halo of our Galaxy. We
consider
 a ``standard'' spherical halo (Bahcall \& Soneira \cite{BS}, Caldwell \& Ostriker \cite{CO})  
\beq 
\rho(r) = \rho_\odot \frac{r_\odot^2+a^2}{r^2+a^2} 
\label{rho} 
\eeq
cut at a distance of 100 kpc, where the density in the solar neighbourhood is
$\rho_\odot \simeq 0.008 M_\odot/\mathrm{pc}^3$ (Flores
\cite{flores}), the core radius $a$ ranges from 2 kpc (Bahcall \&
Soneira \cite{BS}) to 8~kpc (Caldwell \& Ostriker \cite{CO}), and the
distance from the sun to the galactic centre is $r_\odot = 8.5$
kpc. Assuming an isotropic distribution for the transverse velocity
$V_\bot$ of halo objects\footnote{This approximation is sufficient to
  get an order of magnitude.}, the value of $\Gamma_0$ in the
direction of M~31 is:
\beq
\Gamma_0^{\mathrm M~31} \simeq 7\times10^{-6} \; {\mathrm year}^{-1} \;
\frac{<V_\perp>}{\mathrm 200 \; km/s} \; \left[ \frac{0.1 \;
    M_\odot}{M_{\mathrm bd}}\right]^{1/2}\ ,
\eeq
taking into account only lenses of the halo of our galaxy. 
   
The amplification $A$ required for detection depends on the magnitude $m$ 
of the star and on the surface magnitude $\mu$ of the
background at the pixel position. The number of photoelectron/s actually
counted by the CCD on our reference image, from a star of magnitude $m$ is:   
\beq
F_{\mathrm{star}} = F_0 \; 10^{-0.4 m}\ .
\eeq
We measure 
$F_0 = (1.5 \pm 0.1)\ 10^9$ photoelectron/s with the Gunn r filter and
$F_0 = (1.9 \pm 0.1)\ 10^9$ photoelectron/s for the Johnson B filter
(see Eqs.~(\ref{eq:magn}-\ref{eq:magn2}) below, remembering that the gain of the CCD is 9.4). 

To compare with other instruments, note that effective fluxes are related to
photon fluxes ${\cal F}$\ (in $\mathrm{cm}^{-2}\,s^{-1}$)  outside
the atmosphere by:
\beq
F =\frac{\pi\,{\cal F}\,\Delta^2}{4\,\epsilon_{\mathrm{CCD}}\,P}. 
\eeq
Here $\Delta$ is the
diameter of the telescope, $\epsilon_{\mathrm{CCD}}$ is the quantum efficiency
of the CCD camera, and $P$ is a variable loss factor, both atmospheric
and instrumental, which is typically about~3.
 
Neglecting the night sky background (this is justified near the bulge
of M~31), the number of photoelectron/s counted per square arcsecond from the
background is:
\beq
F_{\mathrm{galaxy}} = F_0 \; 10^{-0.4 \mu}
\eeq
Since the light of the galaxy is nothing but the integrated light of all 
stars,
we get the very useful relation: 
\beq
\int 10^{-0.4 m} \phi(m) \mathrm{ d}m = 10^{-0.4 \mu} \label{flum} 
\eeq
where $\phi(m)$ is the luminosity function of the galaxy (here defined as 
the
number of stars of magnitude between $m$ and $m+\mathrm{ d}m$ per 
arcsec$^2$). When the star is microlensed, the signal in a pixel is, if the
exposure time $t_{\mathrm{ exp}}$ remains small compared to $t_E \times u_0$: 
\beq
\mathrm{ Signal} = (A-1)\, F_{\mathrm{ star}} \; f \; t_{\mathrm{ exp}}
\eeq
where the seeing fraction $f$ is the fraction of the star flux that reaches
the pixel. We estimate that our level of noise
is approximately twice the statistical photon fluctuation (see section
\ref{sphotal}): 
\beq
\mathrm{ Noise} = 2\, \left(F_{\mathrm{
      galaxy}}\;\Omega_{\mathrm{pixel}} \; t_{\mathrm{ exp}}\right)^{1/2}
\eeq
where $\Omega_{\mathrm{pixel}}$ is the angular surface of the pixel, in
arcsec$^2$. 
If one wants that the signal to noise ratio be larger than $Q$, then
the lens must approach the line of sight of the lensed star nearer than 
\beq
u(m) =\frac{10^{-0.4\,m}}{10^{-0.2\,\mu}}
\frac{f}{2\,Q}\left[\frac{F_0\,t_{\mathrm{exp}}}{\Omega_{\mathrm{pixel}}}\right]^{1/2},\label{eq:udem}
\eeq
where we have used the fact that, when the amplification is large, 
$A - 1\,\simeq\,1/u$. We have neglected finite size effects, which would
decrease the number of events for small mass lenses ($< 10^{-3}
M_\odot$) in the halo of M~31. The total number of events with a signal
to noise ratio  above Q is then:
\beq
N_{\mathrm{events}} =
t_{\mathrm{obs}}\,\Omega_{\mathrm{tot}}\,\Gamma_0\,\int u(m)\,\phi(m)
\mathrm{d}m,\label{eq:NEV}
\eeq
where $t_{\mathrm{obs}}$ is the total duration of the observation, and
  $\Omega_{\mathrm{tot}}$ is the total solid angle covered. Taking
  into account
  Eqs.~\ref{flum} and ~\ref{eq:udem}, the shape of the luminosity
  function $\phi(m)$ drops out and one
  finally gets: 
\beq
N_{\mathrm{events}} =
t_{\mathrm{obs}}\,\Omega_{\mathrm{tot}}\,\Gamma_0\,10^{-0.2\,\mu}
\frac{f}{2\,Q}\left[\frac{F_0\,t_{\mathrm{exp}}}{\Omega_{\mathrm{pixel}}}\right]^{1/2}\label{eq:NEV2}
\eeq

Using eq. \ref{eq:NEV2} with $Q = 5$, in the conditions of AGAPE
(described below in section 
\ref{observ}), where the total observation period is 190 days, the
total solid angle covered is $8\arcmin\times8\arcmin$, the super-pixel
size is $2.1\arcsec$, and the mean surface magnitude lies
around $\mu = 19$, we expect about 8 events from Milky Way lenses with a
mass of $0.08 M_\odot$. However, this evaluation is an overestimate,
because it only requires that 
one point of the light curve reaches a signal to noise ratio above 5,
disregarding whatever happens at the preceeding and following points.

\paragraph{M~31 lenses} Lenses in M~31 and its halo act on point-like
sources in the same way as those of the Milky Way because, if one
neglects the angular size of the source, the
lensing phenomenon is symmetric between observer and source. 
The contribution of the lenses in 
M~31 cannot be evaluated in the same simple way, for two reasons. i) For
low mass lenses in M~31 or in its halo, the angular Einstein radius is
not much larger than the angular radius of most bright stars, which
can no more be considered as point-like. As a result, the
amplification is limited by finite size effects and
seldom becomes large enough to be detectable. In fact lenses lighter
than $10^{-4} M_\odot$ around M~31 produce nearly no detectable
microlensing. 
ii) On the contrary, for high masses, one expects lenses around M~31
to dominate, because M~31 is roughly twice as massive as the Milky Way,
and because bulge-bulge lensing should be important in the central region we
are looking at (Han \& Gould \cite{HG}). The distribution
of M~31 lenses, and therefore their 
contribution to the lensing rate, strongly depends on the region of
the galaxy one considers. As a matter of fact, this is an advantage,
because (Crotts \cite{crotts})  it 
provides a signature of the lensing phenomenon, and it will allow to
make a map of the distribution of M~31 lenses if one achieves enough
statistics.

\paragraph{Numerical simulation} To give ourselves the possibility: 
i)~to take into account the lenses of M~31,
ii)~to put into our evaluations the real event selection criteria and
to change them,
iii)~to work with the true observation conditions, such as the varying
seeing and the real distribution in time of the observation nights, 
iv)~to play with the distributions, still poorly constrained, of the
lenses and source stars both in the Milky Way and in M~31, 
we have built a Monte-Carlo simulation. 
Typical inputs for the simulation are as follows. The halo of our
galaxy is taken ``standard'' (Eq.~\ref{rho}) with a core radius $a$ of
5 kpc, the halo of M~31 is taken twice as large. An event is called
detected if the light curve shows a series of at least three
consecutive points with a signal to noise ratio above 3 and above 5
for one of these points. With
these assumptions, the number of expected events is about 3 from the
Milky Way halo, and 8 from the M~31 halo. Bulge-bulge lensing in M~31 has not
yet been included in our simulations but, according to Han \& Gould
(\cite{HG}), should contribute as much as lensing by the M~31 halo.
One must,however, emphasise that the number of events one expects depends on
the detailed process of analysis and on the event selection, which are
not settled at this stage.

It is interesting to compare qualitatively the Monte-Carlo simulations
with the  analytic expressions above which, although crude numerically,
show some interesting features.
\begin{enumerate}
\item As can be seen from Eq. \ref{eq:NEV2}, the lensing rate does not
depend on the shape of the luminosity function
$\phi(m)$ of M~31. This is quite welcome since this function is largely unknown 
(except for the brightest resolved stars) and moreover it changes from
the centre to the outskirts of M~31. Our Monte-Carlo simulation indeed
confirms that the rate depends only weakly on the shape $\phi(m)$ 

\item The lensing rate scales with the galactic surface brightness as 
10$^{-0.2\mu}$ as a result of the competition between the number of
source stars and the photon noise. Our Monte-Carlo 
simulation confirms this behaviour. This scaling in $\mu$ is related
to the statistical nature of the fluctuations, which is proportional
to the square root of the number of photons. It is
certainly wrong when the statistical error is very small, then
we know that 
other sources of 
fluctuations, such as the Tonry-Schneider
surface brightness fluctuations (Tonry \& Schneider~\cite{TS}), and
the residuals of the geometric alignment, take over. We take into
account this fact in our 
Monte-Carlo simulations by setting a lower bound on the relative fluctuation.
As we shall see in section~\ref{sphotal}, this bound is not higher
than 0.1\% in our data. This lowest level of fluctuation is of crucial
importance: if we were only able to reach 0.2\%, the expected number
of events would drop by a factor of~3.
\end{enumerate}
\figun
The monte-carlo simulation allows to predict the distribution of
various quantities that characterise microlensing
events. In Fig.~\ref{fig01} the distributions of two time
scales are compared: i) the effective duration of the events
$t_{\mathrm{eff}}$, i.e. the time during which an event is effectively
detected with a signal to noise ratio higher than 3; ii)
twice the  Einstein time $t_E$ (twice because, in 
comparing with the effective time, the
diameter rather than the radius of the Einstein ring is relevant to
the total duration of an event). The two distributions are very
different. The absence of events with an effective duration
$t_{\mathrm{eff}}$ between 100 and 240 days is related to the
distribution of our observation periods: first 60 days in 1994, then
a 240 days gap, and finally 150 days in 1995.

\figdeux
Figure~\ref{fig02}
displays the distributions of the absolute V~magnitude of lensed
stars, and of the amplification at maximum in the conditions of the
real observation. As expected, the stars involved in
detectable microlensing events are giants, and the amplifications are
high, with a mean value of about 13.

In Fig.~\ref{fig03}
\figtrois
are displayed super-pixel light curves of {\em simulated} microlensing
events satisfying our detection criteria, in the real observation conditions.
\section{The experiment \label{observ}}
\subsection{Observational settings}
Data were taken on the 2 metre telescope TBL,
during a 2 months period in 1994 (September 28 to November 24),
 and during 93 nights scattered over 6 months (July to December) in
1995. Observations were only carried out when M~31 was higher than
35$^\circ$ above the horizon. 
\paragraph{Optical device} We observe at the f/25 Cassegrain focus
behind a focal reducer ``ISARD'' that brings the aperture to f/8, in a
$4.5\arcmin\times4\arcmin$ field where the image quality is compatible
with the 0.3\arcsec sampling of the CCD camera.  
\paragraph{Filters} To be able to test achromaticity, we use two well
separated filters: Johnson B and Gunn r. 
\paragraph{CCD Camera}
The camera is a 1024$\times$1024 Tektronix CCD camera. Pixels are 24
$\mu$m wide, which corresponds to an angular size of $0.3\arcsec$. The
effective field covered by ISARD is only 900$\times$780 pixels.  The
chip is thin and its quantum efficiency remains above 70\% in the two
bands we use. The array is very clean with very few bad pixels.
The readout noise is $12\,e^-$ and the gain, or conversion factor is 
$9.4\,e^-/ADU$. 
\paragraph{Exposure time} 20 minutes in red\footnote{Except for the first
exposures in 1994, when ISARD was tuned in a less efficient way.} and
30 minutes in blue. 
\paragraph{Runs} For various reasons, and in particular because the telescope we use is
not dedicated, the focal reducer ISARD must often be
dismounted and remounted. After such an operation, the positions of
the mirrors and the camera are never exactly the same as before.
We call a session between two dismounting-mounting of ISARD a
``run''. Our exposures were taken over a total of 10 runs in our two autumns of
observation, each of which is identified by a letter a,b,c ...

\subsection{Observations} 

As the field of ISARD is small, we were led to cover the M~31
bulge with 6 fields (fields A, B, C, D, E and F of
Fig.~\ref{fig04}). An additional field, Z, centred on the nucleus of 
M~31 was taken at the beginning of each night, 
as a reference to help in the pointing of the telescope.
\figquatre
It turned out that it was impossible to monitor all the fields in
both colours each night. We decided to put a priority on the first four
fields, with an emphasis on red exposures. Blue images, which require
longer exposure times, were less regularly taken. Fields E and F were
poorly sampled. We had altogether 76 nights of good weather over the
two periods of observation. The number of images taken in each
field during the whole survey is summarised in Table~\ref{table:1}.
\begin{table}[ht!]
\caption[]{Number of pictures taken for each field in both
  colours, over the two periods of observation.\label{table:1}}  
\begin{flushleft}
\begin{tabular}{cccccccc}
\hline\noalign{\smallskip}
Field & A  & B  &  C &  D &  E &  F & Z \\
\noalign{\smallskip}\hline\noalign{\smallskip}
Red   & 76 & 66 & 60 & 56 & 40 & 32 & 83\\
Blue  & 32 & 31 & 24 & 19 & 10 &  8 & 32\\
\noalign{\smallskip}\hline
\end{tabular}
\end{flushleft}
\end{table}

\subsection{Pre-processing} 
Raw data were processed at Pic
du Midi during the observation sessions using MIDAS.
Mean bias images have been constructed for each night, from a median
combination of typically 10 frames, and show a good stability.
Mean flat-fields have been made for each run and they correct most of
the  differences between runs. We come back on this point later.

\section{Data reduction \label{treatment}} 
Because observing conditions are never the same for two successive
exposures, three corrections have to be applied to the
images before pixel light curves can be extracted:

\begin{enumerate}
\item A pixel light curve makes sense only if a definite pixel always
covers the same part of the sky on all successive pictures to a very
high degree of precision (within $0.1\arcsec$). This is never the case for raw
data to such an accuracy, and we correct for that by
software. We call the corresponding correction {\em geometric alignment}.

\item Atmospheric conditions are never the same. In particular, the
absorption of light and the sky background change significantly from
one exposure to the other (in particular with the moon). The
corresponding correction is called {\em photometric alignment}.

\item Seeing changes from night to night and this must also
be corrected for. However, when dealing with large enough super-pixels
far from bright stars it can be neglected in a first step.
\end{enumerate} 

\paragraph{Reference image}
To apply geometric and photometric alignment, one must choose a
reference image. We have chosen images taken on October 26
1994, because observing conditions were good and all fields A to F were
available in both colours. 

\subsection{Star detection and seeing\label{seeing}}
\figcinq
To find out a maximum of stellar objects on our pictures, we used an
adapted version of the program PEIDA, developed by one of us within the EROS
collaboration (Ansari \cite{PEIDA}), which is optimised to process quickly a
large number of images.
The main changes we had to implement concern the
small number of resolved stars (around 50 per field)
and the strong gradient of the background, which compelled us to
rethink the star detection. 

This treatment left us with 56 stellar objects on the reference image
of the A field. Each object plus its background was then fitted by a 
two-dimensional Gaussian PSF plus a plane ($9$
parameters altogether). In this way we get the value of the full width at
half maximum (FWHM) for each object. 

The next step was to distinguish the ``real'' stars
from other types of objects such as globular clusters, which would
artificially increase the average seeing of the picture.
We did so using the following discriminating method:  
if on most pictures the  FWHM of an object was
\emph{significantly} above the average, it was removed from the
average estimate, and the process was iterated. After this treatment,
we ended with a  total of 32 ``real'' stars in each of our pictures of
the A field. 

This procedure allowed us to discard a few bad images, where the $\chi^2$
of the PSF fit was poor for most of the 32 stars. We were left with 64
exposures of good quality for the A field, for which the average seeing
for the 1994 runs was $1.5\pm0.4\,\arcsec$, and $1.6\pm0.4\,\arcsec$
for the 1995 runs. 
Fig.~\ref{fig05} shows the evolution with time of the seeing in
1994 and 95, and the distribution of the
seeing for both years combined.

Seeing is highly variable and this is a major problem. As mentioned
earlier, we cope with these seeing variations by working 
with super-pixels $2.1\arcsec$ wide obtained by replacing {\em each}
elementary pixel by the square of $7\times7$ elementary pixel centered
on it. Far from bright stars, this is sufficient for seeings smaller
than $1.8\arcsec$, even if we expect to do better in the future.

\subsection{Geometric alignment}
\figsix 
Geometric alignment involves a two steps procedure:
\begin{enumerate}
\item On the reference and the current images, one detects 
 as many bright stars as possible, 
  one identifies them on the two frames, and one computes the general
  linear transformation in two dimensions, sometimes called the ``Turner
  tranformation'', that corrects for any 
  translation, rotation and scale change between the current and the
  reference images.
\item The Turner transformation of the current image to the reference
image is implemented by linear interpolation. In general, this can
become very complicated 
  as each pixel is not only translated but also scaled and
  rotated. However, rotations and scale transformations are very small
  and, although they are important for the \emph{position} of the
  transformed pixel, the changes they induce on the pixel
  orientation, size and shape may be neglected. 
\end{enumerate}
This geometric alignment is quite successful as can be seen in
Fig.~\ref{fig06}. The dispersion of the differences in star
positions on two images, after
alignment, is of the order of 0.3 pixel, that is $0.1\arcsec$. However
this dispersion is dominated by the uncertainty on the determination
of the position of each star, therefore the precision of the geometric
alignment is\emph{ better} than $0.1\arcsec$  

\figsept

\subsection{Photometric alignment \label{sphotal}} 

In general, photometric alignment is performed assuming that all
differences in instrumental absorption between runs have been removed by the
correction for flat fields. In this case one may assume the existence of a linear
relation (supposing identical seeing) between the
intensity in corresponding pixels of the current and reference images: 
\beq 
F^{\mathrm{reference\ image}}_{\mathrm{pixel}} = a\,
F^{\mathrm{current\ image}}_{\mathrm{pixel}} + b\ . \label{eq:frel} 
\eeq
Here $a$ is the ratio of absorptions (due to variations of the atmospheric
transmission and/or airmass effects) and $b$ the difference of sky
backgrounds (due to moon phases, and/or variations of the
atmospheric diffusion) between the reference and the current image.

The usual way to evaluate $a$ is to compare the total intensities of
corresponding stars on the two pictures. However,
we cannot get in this way a precision better than a few
percent on the factor $a$,  because the photometry 
can be done only on about 50 stars and is difficult on each star,
because they are
faint and the background is very steep. For this reason, we devised an
original global statistical 
approach to tackle the problem, global in the sense that we take into
account all pixels, and not only a few resolved stars. The two methods
give equivalent results, but the statistical approach allows to push the
precision to about 0.5\%. 
\paragraph{Statistical approach}
Assuming relation (\ref{eq:frel}), the variance $\sigma^2$ and the mean
value $<\!\mathrm{image}\!>$ of the histograms of pixel intensities on the two images are
related by: 
\alpheqn
\beqa
\sigma^2_{\mathrm{reference\ image}}&=&a^2\,
\sigma^2_{\mathrm{current\ image}} \label{eq:varela}\\
b&=&<\mathrm{reference}> -\, a <\mathrm{current}>   
.\label{eq:varelb}
\eeqa
\reseteqn
Relations (\ref{eq:varela},\ref{eq:varelb}) are valid only when the main cause of
variance is the gradient of the surface brightness of M~31. The photon noise and
fluctuations due to seeing variations can in principle invalidate
equation (\ref{eq:varela}). However, in our case, the luminosity
gradient of the bulge of M~31 largely supersedes all other causes of variations.
The efficiency of this procedure is illustrated in Fig.~\ref{fig07}:
pixel histograms, for four pictures, that look very different before
treatment coincide down to small structures after photometric
alignment, using only the two parameters $a$~and~$b$.
\fighuit
\subsection{Filtering out of large spatial scale variations}
\paragraph{Reflected light}
After photometric alignment, there remains a slight
gradient in the difference between two images of different
runs. This is particularly obvious between runs c~and~d, when we had to
take ISARD down and tune its mirrors. This resulted in a substantial
gain of luminosity but introduced a significant gradient between
images of runs c~and~d (Fig.~\ref{fig08}).

We think that this residual gradient is due
to reflected light for the following reasons. i) It is not cured by the
usual debiassing and flat-fielding procedures. ii) Its shape depends
on the field but seems constant for each field in a given run. iii)
Its intensity seems proportional to the overall luminous
intensity. 

\paragraph{Median background image}
To cope with the problem, we construct for each frame a background
image where the stars are 
removed using a median filter. We take a $41\times41$ window for
the median filter, that is with a surface much larger than that of the
largest seeing disk, therefore all stars but the very
brightest completely disappear. We then subtract from each frame its
background image and add that of the reference frame. 

\paragraph{High spatial passband filter}This procedure filters out
variations of low spatial 
frequencies: it insures that, relative to the reference image, all
variations on 
scales larger than 40 pixels are very strongly suppressed whereas
variations on scales smaller than 20 pixels are fully preserved. The
only remaining 
differences between images come either 
from short scale fluctuations (seeing variations around stars and
around surface brightness fluctuations, or photon noise) or from
varying stellar objects. 

\paragraph{Residual gradient and the alignment coefficient $a$}
Because of this residual gradient,
 the sky backgrounds of two images do not
stricly satisfy Eq.~(\ref{eq:frel}). This introduces a
systematic error on $a$ when comparing different runs. This error,
however, remains smaller than the error arising 
from matching resolved stars. As all
images have been brought to have the same median background, the error
on $a$ only affects the difference of the super-pixel intensity with this background and not the
total super-pixel intensity. In other word, the
systematic uncertainty on $a$ does not alter our ability to
detect variations, but it limits our precision on the time evolution
of a variation, once detected.  

\paragraph{}
The pixel stability in time achieved after the processing presented
above is described in section \ref{results}

\subsection{Absolute photometric calibration \label{calgun}}

Absolute photometric calibration is, strictly speaking, not necessary
for microlensing searches which rely solely on the detection of
\emph{relative} luminosity variations in time. Nonetheless, to study
the nature of the variable objects we detect, it is necessary to know
their absolute magnitude. 

We took images of the Palomar-Green PG1657+078 calibration field from
Green \emph{et al.} (1986) on 28 July 1995 (calibration day).
To determine the flux of reference stars reported in
Landolt (\cite{landolt}) \textit{UBVRI} photoelectric observations, we used the same 
procedure as for the study of seeing (see section~\ref{seeing}) except
that the fit with a gaussian plus a plane is used only to
determine the plane that fits the background, the flux of the star
is then obtained by subtracting the estimated background to the
observed total flux under the star. The photometry obtained in this
way turns out to be much more stable among different images.
The colour equations for the Johnson
R and B magnitudes, denoted $m_R$ and $m_B$, are:
\beq
\begin{array}{cll}
m_R & = & \alpha + r + \beta (b-r)\\
m_B - m_R & = & \gamma + \delta (b-r),\label{eq:magn}
\end{array}
\eeq
where $r$ and $b$ are the instrumental magnitudes with the Gunn r and
Johnson B filters:\\
$
r \mbox{ or }b = -2.5 \log F_{\mathrm{star}}^{\mathrm{red\ or\ blue}}
$.\\
We find, using a $\chi^2$ minimisation: 
\beq
\begin{array}{llllll}
\alpha & = & 21.29\pm 0.02,\ & \beta  & = & 0.05 \pm 0.02 \\
\gamma & = & 0.23 \pm 0.03,\ & \delta & = & 0.89 \pm 0.03.
\end{array} \label{colour}
\eeq
We then have to transform our results for $\alpha$ to the reference
day where atmospheric absorption was different. The final value is: 
\beq 
\alpha = 20.50 \pm 0.05\label{eq:magn2}
\eeq
and the other coefficients are not affected.

\section{
Light curves\label{results}}
The pixel method relies on the inspection of pixel light curves.
\figneuf
Light curves are graphs of the variation of pixel intensities.
Elementary pixels are small ($0.3\arcsec$), which is very useful to get a good
geometric alignment. However, elementary pixels undergo strong
fluctuations due to seeing variations that hamper detection of truly
variable stellar objects. For this reason we replace 
{\em each} pixel by a super-pixel, as explained in section
\ref{method}. A convenient size for the super-pixel, in vue of the
average seeing of $1.5\arcsec$, turns out to be $2.1\arcsec$, wich
corresponds to super-pixels built with $7\times 7$ elementary pixels. 

\figdix
Using super-pixels provides a substantial gain in stability.
Figure~\ref{fig09} shows maps of the relative fluctuation along the
light curve of elementary $0.3\arcsec$ wide pixels, and of 
$2.1\arcsec$ wide $7\times 7$ super-pixels of field A (Notice that there are as
many super-pixels as elementary pixels). On elementary
pixels, the dispersion is below 1\% on most of the field. For
super-pixels, the dispersion drops
down to 0.3\% in average and even reaches a level
below 0.1\% in the most stable regions, as announced earlier. It
remains everywhere around twice the photon noise.   

To compare in more detail the super-pixel fluctuation to the photon
noise, we have computed along the light curve of each super-pixel the 
$\chi^2$ of the difference between the intensity on the current image and
its average in time. 
In Fig.~\ref{fig10}, we display the distribution of this $\chi^2$
for the super-pixels of field A, using two different seeing selections.
The error $\sigma$ 
entering the $\chi^2$ is chosen in such a way that the maximum of the
distribution of the $\chi^2$ coincides with that of the ideal Poisson 
law. This is achieved for $\sigma \simeq 1.7 \sigma_\gamma$ where $\sigma_\gamma$ is
the statistical photon noise. The true distribution shows non-poissonian
tails. Clearly there are non-statistical contributions to the
fluctuations and a comparison between Fig. \ref{fig10}a and
Fig.~\ref{fig10}b shows that they are largely due to seeing 
variations. Further work is in progress to cope with the latter.
This non poissonian behaviour is also responsible for the fact that, in
going from pixels to super-pixels, one gains less than the factor 7
expected if fluctuations were of pure statistical origine.

We have made the same study replacing super-pixels by a PSF weighted
average. The fluctuation is twice larger than with
super-pixels, and the tails due to seeing variations in the $\chi^2$
distribution are much larger.

Figure~\ref{fig11} illustrates the considerations
above with the light curve of a 
stable super-pixel, keeping only the frames with seeing between $1.1\arcsec$ and
$1.8\arcsec$. Super-pixel intensities are in ADU/s (1 ADU/s on a 2.1$\arcsec$
super-pixel corresponds to a surface magnitude $\mu_R=22.1$). The
R.M.S fluctuation along 
\figonze
\figdouze
the light curve is 0.045 ADU/s, to be compared with the average photon
noise which is around 0.04 ADU/s. If one keeps all points,
irrespective of the seeing, the RMS
fluctuation becomes 0.065 ADU/s. The error bars correspond to 1.7
$\sigma_\gamma$, that is around 0.07 ADU/s in average.\\

With this level of stability, we are able to clearly see variations at
the level of a few percent as is apparent from Fig.~\ref{fig12}.
Let us stress the following features of this figure.
\begin{enumerate}
\item This light curve shows two clear variations, it is a variable
  star, not a microlensing.  
\item On graph (b), only points corresponding to a seeing between
  $1.1\arcsec$ and $1.8\arcsec$ have been retained and the light curve appears
  much smoother than on graph (a).
\item After seeing selection (graph (b)), the first variation can
clearly be seen, because of its coherence in time, although it is only
about 0.5 ADU/s, that is about 5 times the average error bar in this period.
\end{enumerate}
We see that the the selection criteria we have introduced in our
Monte-Carlo simulation in section~\ref{NOV} (3 points above~$3\sigma$
and one of them above $5\sigma$ ) are indeed realistic. 
However our present thresholds are much higher, because these criteria
would be sufficient if microlensing were the 
only possible source of variations. This is of course not the case and
variable stars are far more numerous. If we used only the criteria
of our simulation, we would be swamped by variable objects. Therefore,
to isolate 
microlensing events we have to build filters which reject most of the
variable objects but not the microlensing events satisfying our
criteria. There are many conditions that can be added, such as:
\begin{itemize}
\item the usual conditions of unicity, symmetry and achromaticity
\item the quality of fits by a Paczy\'nski curve 
\item limits on the duration of events expected from MACHO's with
reasonable masses compared with what is expected from simulations (see
figure~\ref{fig01}).
\end{itemize}
We are working on that. We will be in a better position after the
30 observation nights we shall have in autumn 1996. Although these nights will
be too few and too scattered to allow detection of new events, they
will allow to constrain efficiently fits of events that occured in
1994 and 1995

Even events that overshoot by far our criteria would have been
extremely difficult to detect by monitoring resolved stars.
\figtreize
This is illustrated in Fig.~\ref{fig13}.
The two dimensional surface plots (a) and (b) map the intensities of
elementary pixels around the centre of a detected variation. Plot (a)
corresponds to the minimum of the light curve and plot~(b) to the
maximum. Most structures appear similar on the two
plots, which means that they correspond to real structures of M~31.
They are the surface brightness fluctuations of Tonry \& Schneider
(\cite{TS}). At the centre however, a tiny
bump, barely visible on graph (a), has grown into a clear PSF-shaped
peak on (b). This tells us that we are really looking at a varying
stellar object, barely detectable as a resolved star. 

Variable stars are interesting in their own right.
Numerous variable objects such as the preceding ones have been
detected, but we are only beginning to analyse their nature.
\figquatorze
Figure~\ref{fig14} shows the light curves of two objects, one
of which is probably a cepheid, and the other a nova. We have a  host of other
cepheid candidates and five novae with peak magnitude and rate of
decrease similar to the 
one shown on Fig.~\ref{fig14}, and very similar to the
M~31 novae quoted in Hodge \cite{hodge}.

We also see variations compatible with microlensing (about 20). However at this
stage, we are not in a position to claim that we have seen microlensing
events for several reasons. First, our lever arm in time is not
sufficient to be sure that the variations do not repeat, and even in
some cases, to
be sure that events are really symetric. The situation will improve
with the 30 nights we expect in autumn 1996. Second, we have not yet
analyzed the blue light curves, therefore we cannot yet test
achromaticity. Figure~\ref{fig15} shows one of these light curves. 
\figquinze
The Paczi\'nsky curve on Fig.~\ref{fig15} corresponds to a star of
absolute magnitude M=-2 amplified by a factor 6 at maximum and with an
Einstein 
time scale $t_E = 65$ days. These number are not well
determined because of a parameter degeneracy for high
amplification events (see for instance Gould \cite{gould}), which is
the case of most events we can detect. A time scale and  a maximum amplification twice as
large associated with a star twice fainter would fit just as well. 
However the time scale cannot be much shorter, because the star should
be brighter and would be seen more clearly before the lensing begins. 
The effective time $t_{\mathrm{eff}}$ is 19 days if one measures it
between real points of observation where the signal to noise ratio is
higher than 3, and 40 days if one refers to the time
during which the Paczi\'nsky curve remains at 3 $\sigma$ above the
background. This effective time is a powerful mean to eliminate fake
microlensing events: our simulation tells us 
that $t_{\mathrm{eff}}$ should be smaller than 60 days for lenses with 
masses around 0.08 $M_\odot$. The numerous time gaps we have
in the observations make it difficult to find short events, and it is
important to our approach to have as few gaps as possible in the
time sampling.

\section{Conclusion}
On the basis of data taken during two autums at the 2 metre telescope
Bernard Lyot at Pic du Midi, AGAPE has proven that the pixel method
works. Super-pixels, taken $2.1 \arcsec \times 2.1 \arcsec$ large to
minimise the effect of seeing variations, have a level of fluctuation
not larger than 1.7 times the photon noise. On the brightest
super-pixels this fluctuation is not more than 0.1\% of the photon
background. With such a stability, our simulations predict that we
should see around 10 events in the direction of M~31 for lenses with
$0.08 M_\odot$ masses, an event beeing called detectable if its light
curve remains 3 
$\sigma$ above the background for at least 3 consecutive points and
reaches 5 $\sigma$ at one of them. Such variations are clearly
detectable from their time coherence, even if more work is needed to
separate microlensing events from other kind of variations. We are
already detecting hundreds of variable stellar object in M~31, in
particular cepheids and novae. They are currently beeing analysed. 

To exploit the full power of our method, we are exploring the
possibilities of launching an 
observation with a wide field camera on an instrument were we could
get a very regular and short time sampling, and a lever arm of several
years. We would then be able to make a map of the halo of M~31, which
would be of considerable interest for halo model builders.

\acknowledgements{
We wish to thank F. Colas, D. Gillieron, and A. Gould for useful
discussions and suggestions}


\begin{thebibliography}{}

\bibitem[1995]{DUO} Alard, C., Mao, S., Guibert, J. 1995, Object
  DUO 2: A New Binary Lens Candidate, to appear in A\&A Letters 

\bibitem[1996]{ALARD} Alard, C. 1996, communication at the 2nd
    International Workshop on Gravitational Microlensing Surveys,
    Orsay, France

\bibitem[1993]{MACHOa} Alcock, Ch. et al. 1993, Nat, 365, 621

\bibitem[1995a]{MACHOb} Alcock, Ch. et al. 1995a, Phys. Rev. Lett., 74, 2867
 
\bibitem[1995b]{MACHOBULGE} Alcock, Ch. et al. 1995b, {\em The
    MACHO Project: 45 Candidate Microlensing Events from the First Year
    Galactic Bulge Data}, submitted to ApJ
 
\bibitem[1996]{FALLARD} Allard, F. et al. 1996, {Synthetic spectra
  and mass determination of the brown dwarf Gl229b}, to appear in ApJ letters

\bibitem[1994]{PEIDA} Ansari, R. 1994, {\em Une m{\'e}thode
  reconstruction photom{\'e}trique pour l'exp{\'e}rience EROS}, Laboratoire
de l'Acc{\'e}l{\'e}rateur Lin{\'e}aire d'Orsay, France, report LAL 94-10

\bibitem[1995a]{EROSb} Ansari, R. et al. 1995a, {\em Observational
    limits on the contribution of sub-stellar and stellar objects to
    the galactic halo}, to appear in A\&A

\bibitem[1995b]{ROME} Ansari R. et al. 1995b, {\em AGAPE, a
    microlensing search in the direction of M~31: status report}, to
  appear in the proceedings of ths workshop ``The Dark Side of The
  Universe'' at University Roma 2, edited by Bernabei R., preprint LPC 96 04/conf

\bibitem[1993]{EROSa} Aubourg, E. et al. 1993, Nat, 365, 623.

\bibitem[1980]{BS} Bahcall, J.N., Soneira, R.M. 1980, ApJS, 44, 73 

\bibitem[1992]{BBGK1} Baillon, P., Bouquet, A., Giraud-H{\'e}raud, Y., Kaplan,
  J. 1992. {\em Search for dark matter as brown dwarves by looking at
  Andromeda (M~31)}, Proceedings of the first Palaiseau Workshop, Fleury,
  P., Vacanti G. editors, Edition Fronti{\`e}res, 151

\bibitem[1993]{BBGK2} Baillon, P., Bouquet, A., Giraud-H{\'e}raud, Y.,
  Kaplan, J. 1993, A\&A, 277, 1 

\bibitem[1996]{BMG} Basri, G., Marcy, G.W., Graham, J.R. 1996, ApJ, 458,600

\bibitem[1984]{MOND} Beckenstein, J., Milgrom, M. 1984, ApJ, 286, 7

\bibitem[1996]{BENN} Bennett, D. 1996, communication at the 2nd
    International Workshop on Gravitational Microlensing Surveys,
    Orsay, France

\bibitem[1981]{CO} Caldwell, J.A.R., Ostriker, J.P. 1981, ApJ, 251, 61

\bibitem[1996]{BBNS} Cardall, C.Y., Fuller G.M. 1996, astro-ph/9603071, ApJ, submitted 

\bibitem[1984]{CBA} Carr, B.J., Bond, J.R., Arnett W.D. 1984, ApJ,
  277, 445 

\bibitem[1995]{colley} Colley, W.N. 1995,
AJ, 109, 440 

\bibitem[1996]{COUCH} Couchot, F. 1996, communication at the 2nd
    International Workshop on Gravitational Microlensing Surveys,
    Orsay, France

\bibitem[1992]{crotts} Crotts, A.P.S. 1992, ApJ, 399, L43

\bibitem[1991]{DJM2}
De R{\'u}jula, A., Jetzer, P., Mass{\'o}, E. 1991, MNRAS, 250, 348 

\bibitem[1992]{EVAP} De R{\'u}jula, A., Jetzer, P., Mass{\'o}, E. 1992, A\&A,
  254, 99 

\bibitem[1995]{dolgov} Dolgov A., D. 1995, Lectures at ITEP Winter
School, Zvenigorod, Russia, astro-ph/9509057, to appear in Surveys of
High Energy Physics

\bibitem[1979]{FG} Faber, S.M., Gallagher, J.S. 1979, ARA\&A, 17, 135

\bibitem[1988]{flores} Flores, R. 1988, Phys. Lett. B215, 73

\bibitem[1996]{gondolo} Gondolo, P. 1996, communication at the 2nd
    International Workshop on Gravitational Microlensing Surveys,
    Orsay, France

\bibitem[1995]{gould} Gould, A. 1995, {\em Theory of pixel
    lensing}, Ohio State University preprint 

\bibitem[1991]{griest} Griest, K. 1991, ApJ, 366, 412  

\bibitem[1995]{griest2} Griest, K. 1995, {\em The nature of the dark
  matter}, to appear in the proceedings of the International School of
Physics ``Enrico Fermi'' course ``Dark atter in the universe'',
Varenna, July 1995, astro-ph/9510089

\bibitem[1996]{HG} Han, C. Gould, A. 1996,
{\em Galactic versus Extragalactic Pixel Lensing Events toward M~31},
Ohio State University preprint, to appear in ApJ

\bibitem[1979]{HR} Harris, W.E., Racine, R. 1979 ARA\&A, 17, 241

\bibitem[1992]{hodge} Hodge, P. 1992, {\em The Andromeda Galaxy},
  Kluwer Academic Publishers, Astrophysics and Space Science Library,
  vol. 176. 

\bibitem[1994]{jetzer94} Jetzer, P. 1994, A\&A, 296, 426 

\bibitem[1991]{KC} Kerins, E.J., Carr, B.J. 1991, MNRAS, 266, 775

\bibitem[1987]{KK} Kormandy, J., Knapp, G.R. 1987, {\em
    Proceedings of the IAU Symposium 117: Dark matter in the
    Universe}, Reidel  

\bibitem[1992]{landolt} Landolt, A.U. 1992, AJ, 104, 340 

\bibitem[1996]{MRZ} Mart{\'\i}n, E.L., Rebolo, R., Zapatero Ozorio,
M.R. 1996, astro-ph/9604080, to appear in ApJ

\bibitem[1995]{THAL} Melchior, A.L. 1995, P.H.D. Thesis, University
  Paris 6

\bibitem[1996]{MILS} Milsztajn, A. 1996, communication at the 2nd
    International Workshop on Gravitational Microlensing Surveys,
    Orsay, France

\bibitem[1995]{NAKA} Nakajima, T. et al. 1995, Nat, 378, 463

\bibitem[1932]{OORT} Oort, J.H. 1932, Bull. Astron. Inst. Netherlands, 6, 249

\bibitem[1973]{OP} Ostriker, J.P., Peebles, P.J.E. 1974, ApJ, 186, 467  

\bibitem[1974]{OPY} Ostriker, J.P., Peebles, P.J.E., Yahil, A. 1974,
  ApJ, 193, L1 

\bibitem[1986]{pacz} Paczy\'nski, B. 1986, ApJ, 304, 1

\bibitem[1994]{PCM} Pfenniger, D., Combes, F., Martinet, L. 1994,
  A\&A, 285, 79, 83 

\bibitem[1995]{RZM} Rebolo, R., Zapaterio Ozorio, M.R., Mart{\'\i}n,
E.L., 1995, Nat, 377, 129

\bibitem[1995]{SACKETT} Sackett, P.D., 1995, `` The distribution
  of dark mass in galaxies'', {\em Proceedings of the IAU
  Symposium 173: Gravitational Lensing}, eds. Kochanek C. and Hewitt J.

\bibitem[1994]{SHP} Stauffer, J.R., Hamilton, D., Probst, R.G. 1994,
AJ, 108, 155

\bibitem[1996]{SUTH} Sutherland, W. 1996, communication at the 2nd
    International Workshop on Gravitational Microlensing Surveys,
    Orsay, France

\bibitem[1994]{TAC} Tomaney, A., Crotts, A. 1994, BAAS,
185, \# 17.01 

\bibitem[1996]{tomaney} Tomaney, A. 1996, communication at the 2nd
    International Workshop on Gravitational Microlensing Surveys,
    Orsay, France

\bibitem[1988]{TS} Tonry, J., Schneider D.P., 1988, ApJ, 96, 807

\bibitem[1987]{trimble} Trimble, V. 1987, ARA\&A, 25, 425

\bibitem[1993]{OGLEa} Udalski, A. et al. 1993, Acta Astron., 43, 289 

\bibitem[1994]{OGLEb} Udalski, A. et al. 1994, Acta Astron., 44, 165

\bibitem[1996]{ZRM} Zapatero Ozorio, M.R., Rebolo, R., Mart{\'\i}n,
E.L. 1996, astro-ph/9604079, to appear in A\&A

\bibitem[1988]{ZB} Zuckerman, B., Becklin, E. 1988, Nature, 336, 656

\bibitem[1933]{zwicky} Zwicky, F. 1933, Helv. Phys. Acta, 6, 110


\end{thebibliography}
\end{document}